\documentclass[11pt,draftclsnofoot, onecolumn]{IEEEtran}
\usepackage{setspace}
\usepackage{algorithm}
\usepackage{algorithmic}
\usepackage{multirow}
\usepackage{dcolumn}
\usepackage{amsmath}
\usepackage{flushend}
 \usepackage{cite}
 \usepackage{color}
\usepackage{amssymb}
\usepackage{psfig}
\usepackage{changebar}
\usepackage{graphicx}
\usepackage{subfigure}
\usepackage{multirow}
\usepackage{array}
\usepackage{enumerate}
\usepackage{balance}
\usepackage{cases}
\usepackage{mathrsfs,amssymb}
\usepackage{stfloats}
\usepackage{url}
\usepackage{slashbox}

\newtheorem{theorem}{Theorem}[section]
\newtheorem{lemma}[theorem]{Lemma}

\newtheorem{corollary}[theorem]{Corollary}
\newtheorem{defn}{Definition}[section]

\newtheorem{cond}{Condition}[section]

\newcommand{\qed}{\nobreak \ifvmode \relax \else
      \ifdim\lastskip<1.5em \hskip-\lastskip
      \hskip1.5em plus0em minus0.5em \fi \nobreak
      \vrule height0.75em width0.5em depth0.25em\fi}
\def\beq{\begin{equation}}
\def\eeq{\end{equation}}
\def\b0a{{\bf 0}}
\def\b1a{{\bf 1}}
\def\bPhi{{\bf \Phi}}
\def\bPsi{{\bf \Psi}}

\def\be{{\bf e}}

\def\br{{\bf r}}
\def\bs{{\bf s}}

\def\bu{{\bf u}}

\def\bw{{\bf w}}
\def\bx{{\bf x}}
\def\by{{\bf y}}
\def\bz{{\bf z}}

\def\bA{{\bf A}}

\def\bD{{\bf D}}

\def\bI{{\bf I}}

\def\bS{{\bf S}}

\def\bU{{\bf U}}

\def\bX{{\bf X}}

\usepackage{cont}

\begin{document}
\title{ Multi-Resolution Compressed Sensing Reconstruction via Approximate Message Passing}
%
\author{Xing Wang, and Jie Liang
\thanks{This work was supported by the Natural Sciences and Engineering Research Council (NSERC) of Canada under grant RGPIN312262, STPGP447223, and RGPAS478109.}
\thanks{The authors are with the School of Engineering Science, Simon Fraser University, Burnaby, BC, Canada. Email: \{xingw, jiel\}@sfu.ca. Corresponding author: J. Liang.}
}
\markboth{IEEE Trans. Computational Imaging}{Wang and Liang: Multi-Resolution Compressed Sensing Reconstruction via Approximate Message Passing}

\maketitle
\IEEEpeerreviewmaketitle
\begin{abstract}
In this paper, we consider the problem of multi-resolution compressed sensing (MR-CS) reconstruction, which has received little attention in the literature. Instead of always reconstructing the signal at the original high resolution (HR), we enable the reconstruction of a low-resolution (LR) signal when there are not enough CS samples to recover a HR signal. We propose an approximate message passing (AMP)-based framework dubbed MR-AMP, and derive its state evolution, phase transition, and noise sensitivity, which show that in addition to reduced complexity, our method can recover a LR signal with bounded noise sensitivity even when the noise sensitivity of the conventional HR reconstruction is unbounded. We then apply the MR-AMP to image reconstruction using either soft-thresholding or total variation denoiser, and develop three pairs of up-/down-sampling operators in transform or spatial domain. The performance of the proposed scheme is demonstrated by both 1D synthetic data and 2D images.

\end{abstract}


\section{Introduction}
\label{sec_intro}

Recently compressed sensing (CS) has been studied extensively as an efficient way of acquiring and reconstructing sparse signals \cite{CS}. Many CS reconstruction algorithms have been developed, e.g., convex optimization \cite{RIP}, greedy method \cite{Coherence}, iterative thresholding \cite{IST}, and approximate message passing (AMP) \cite{AMP,PTCPrediction,DCS-AMP,XingJ,SIAMP,DAMP}.

The AMP is a particularly attractive framework, due to its near-optimal reconstruction performance, low complexity, and the capability of predicting its performance from its state evolution. This leads to the discovery of the phase transition property of the AMP, which states that when the sampling rate of a sparse signal is below a threshold defined by a phase transition curve (PTC) \cite{AMP,PTCPrediction,DAMP,AMPoriginal},  the CS algorithm will fail to recover the signal with high probability even if there is no sampling noise. In the noisy case, the noise sensitivity is unbounded, where the noise sensitivity is the minimax mean squared error (MSE) of the reconstruction. This is analogous to the rate-distortion bound in information theory. Therefore, in applications that a large amount of CS samples need to be transmitted to a receiver, the receiver has to wait until it receives enough samples before it can recover the signal. This could incur undesired delays.

This paper is motivated by the following fundamental question: if in the case above we are allowed to reconstruct low-resolution (LR) previews instead of the original high resolution (HR) signal, can we recover high-quality LR signals so that we can enlarge the feasible operating region of the system? We call this framework CS with multi-resolution reconstructions, or MR-CS for short. It opens up many questions. For example, how to design the sampling and reconstruction algorithms? What is the highest resolution that can be reconstructed at each sampling rate? What are the expressions of the phase transition curves for different LR reconstructions? In addition, a straightforward approach is to first reconstruct a HR signal using existing reconstruction methods, and then downsample it. Therefore another question is how much gain we can get over this simple method. Note that a carefully designed LR reconstruction algorithm should at least have lower complexity than this simple method, because it could reconstruct the LR signal directly.

Although the need for multi-resolution (MR) or scalable reconstruction has been well recognized in multimedia transmission, leading to the development of standards such as JPEG 2000 and H.264/SVC \cite{J2K,SVC},
the problem has received little attention in CS. The schemes that are most relevant to ours are \cite{Bell,STONE,ICIP14}.
 In \cite{Bell}, some rules are proposed to design efficient up-/down-sampling matrices for MR reconstruction, and the number of nonzero entries of the LR image in transform domain is shown to be no larger than that of the HR image. Therefore the required sampling rate for stable LR image reconstruction is less than that of the HR reconstruction. However, the analysis in \cite{Bell} is qualitative, and only some loose bounds are provided. Moreover, the impact of the MR design on the quality of the measurement matrix is not studied, which can be measured by, e.g., restricted isometry property (RIP) constant \cite{RIP} and mutual coherence \cite{Coherence}. Besides, only the noiseless case is considered in \cite{Bell}.

A similar problem to ours is studied in \cite{ICIP14}, where two solutions are proposed. In the first method, the sampling matrix is designed to have non-uniform sampling, which is quite restrictive, since the matrix should be redesigned whenever a new result with different resolution is needed. The second method modifies the sampled data of the HR image to be close to the data acquired directly from the target LR image. Although it works empirically, there is no theoretical guarantee. In addition, although it is mentioned in \cite{ICIP14} that the CS sampling rate for the LR reconstruction is increased, the change of the sparsity rate is not considered. Moreover, the complexity of this approach is even higher than reconstructing the HR image directly. We will show in this paper that the second solution in \cite{ICIP14} is a special case of our proposed MR-AMP framework.

Recently, a special two-resolution CS reconstruction scheme is proposed in \cite{STONE}, where the sampling matrix is designed such that a LR reconstruction can be obtained by direct matrix inversion.


The MR concept has also been used in some CS schemes such as \cite{MRCR,MMSP,KMRCS,MSCS,NCSUAMP} with different purposes from ours. In \cite{MRCR}, Bayesian CS is used to detect the primary user in cognitive radio. It first performs the detection in LR, and then refines the signal around the detected primary user spectrum. In \cite{MMSP}, a CS-based two-layer scalable image coding is proposed, where the encoder employs two measurement matrices with different sizes, and inter-layer prediction is used to reduce the bit rate. In \cite{KMRCS}, the authors extended the Kronecker CS \cite{KCS} to MR measurements, such that the sensing is performed on the LR image, and the goal is to recover the HR signal from LR measurements. In \cite{MSCS}, a multiscale framework is proposed for CS of videos. The motion vectors are estimated at different resolutions and served as the input to higher resolution frame recovery. The sensing is applied to different resolutions of the same frame. In our proposed framework, the sensing is only performed on the original HR image. Therefore, the framework in \cite{MSCS} is more like source coding, but not sensing and coding simultaneously. In \cite{NCSUAMP}, the authors use advanced denoising methods in the multiscale wavelet domain to improve the performance of AMP reconstruction, similar to \cite{DAMP}. However, the reconstruction still has the same resolution as the source.

In this paper, we develop a general theory for MR-CS reconstruction, and propose a MR-AMP algorithm to reconstruct a LR signal if the sampling rate is too low. Our method does not impose any constraint on the measurement matrix. Therefore it enables more LR reconstruction choices. Also, theoretical analysis can still be obtained. Instead of having only one phase transition curve (PTC), we obtain a family of PTCs that specify the sampling rate thresholds to get bounded noise sensitivity with different resolutions. Moreover, the noise sensitivity is derived explicitly. The performance of the proposed scheme is verified using both synthetic data and natural images.

The rest of this paper is structured as follows: Sec. \ref{sec_bkgd} presents the mathematical model of MR-CS problem and provides the necessary conditions that the MR up/down-sampling matrices should satisfy. Sec. \ref{sec_MRAMP} is devoted to MR-AMP algorithm and its updating rule. Sec. \ref{PTC} establishes the theoretical analysis of MR-AMP. Sec. \ref{MRImage} discusses the application of MR-AMP in images, and develops three sets of up/down-sampling matrices. Sec. \ref{sec_experiments} presents simulation results, validates the state evolution of MR-AMP, and gives guidelines on tuning the parameters of the algorithm. It also compares the performance of MR-AMP with the original HR-AMP with different denoisers, in terms of reconstruction quality and algorithm complexity. Some preliminary results of this paper are reported in \cite{XingICIP}.

\section{Formulation and Conditions of MR-CS Reconstruction}
\label{sec_bkgd}

The goal of the classical CS is to recover a ${n_1} \times 1$ vector $\bx$ from a $m \times 1$ noisy measurement $\by$ with $m < n_1$, {\it i.e.},
\begin{equation}
\label{CSeq}
\by = \bA \bx+\bw.
\end{equation}
In this paper, entries of the $m \times {n_1}$ measurement matrix $\bA$ are i.i.d. Gaussian with zero mean and a variance of $1/m$, denoted by $\mathcal{N}(0, 1/m)$. Each entry of the noise vector $\bw$ also follows i.i.d. Gaussian distribution with zero mean and a variance of $\sigma _w^2$. The CS undersampling ratio is defined as ${\delta_1} = m/{n_1}$.

Since the system is underdetermined, it cannot be solved without exploiting the special structure of $\bx$. Some examples of structured signals are given in \cite{PTCPrediction}, including simple sparse signals, block sparse signals, mostly constant non-decreasing signals, and piecewise constant signals. Following the notations in \cite{PTCPrediction}, the family of probability distributions for a particular type of structured signals over ${\mathcal{R}^{n_1}}$ is denoted as ${\mathcal{F}_{n_1, {\varepsilon_1} }}$, where ${\varepsilon _1} \le 1$ is a constant sparsity ratio, and the expected amount of useful structured information in the signals is at most ${k_1}={n_1}{\varepsilon_1}$. The definition of the useful structured information depends on the nature of the structure. Let $\upsilon_{n_1}$ denote a distribution in ${\mathcal{F}_{n_1, \varepsilon_1}}$, and $\bx$ be a signal with distribution $\upsilon_{n_1}$. In this paper, we focus on the following two families of structured sparsity.

\begin{defn}
The family of distributions that generates simple sparse signals is defined as  (Eq. (1.2) in \cite{PTCPrediction})
\begin{equation}
\label{SimpleSparse}
{\mathcal{F}^{SS}_{{n_1},{\varepsilon_1}}} \equiv \left\{ {\upsilon _{n_1}:
{\mathbb{E}_{{\upsilon _{n_1}}}} \left\{ {{{\left\| \bx \right\|}_0}} \right\} \leqslant {n_1}{\varepsilon_1} } \right\},
\end{equation}
where the $\ell_0$ norm ${\left\| \bx \right\|_0}$ denotes the number of nonzero entries of vector $\bx$. Therefore, the expected number of non-zero entries of signals in this family is at most ${n_1}{\varepsilon _1}$.
\end{defn}

\begin{defn}
The family distributions that generates piecewise constant signals is defined as  (Sec. V in \cite{PTCPrediction})
\begin{equation}
\label{Piecewise}
\begin{split}
&{\mathcal{F}^{PC}_{{n_1},{\varepsilon_1}}} \equiv \\
&\left\{ {{\upsilon _{n_1}} : {\mathbb{E}_{{\upsilon _{n_1}}}}
\left\{ {\# \left\{ {t \in \left[1, {{n_1} - 1} \right]:{x_{t + 1}} \ne {x_t}} \right\}} \right\} \leqslant {n_1}{\varepsilon_1} } \right\},
\end{split}
\end{equation}
\end{defn}
where $\#\{\cdot\}$ denotes the number of times the condition in the operator is true. Therefore, the expected number of change points within signals of this family is at most ${n_1}{\varepsilon _1}$.

In the proposed MR-CS reconstruction framework, instead of always recovering the signal with the original resolution $n_1$, we allow the reconstruction of various lower resolution signals $n_d$ ($n_d < n_1$) when the number of available CS samples is too small.

The MR downsampling factor is defined as
\begin{equation}
d = n_1 / n_d.
\end{equation}

Note that this MR downsampling factor should not be confused with the CS undersampling ratio $\delta_1 = m / n_1$. In this paper, we are interested in the case $m < {n_d}$, {\it i.e.}, the recovery of the LR signal is still an underdetermined CS problem. The equivalent CS undersampling ratio for the LR reconstruction is ${\delta _d} = m/{n_d} = d{\delta _1} > \delta_1$. Let $k_d$ be the expected amount of useful information contained in the LR signal. The expected sparsity ratio of the LR signal is $\varepsilon_d = k_d / n_d$. We also define another factor ${\rho _d} = {\varepsilon _d}/{\delta _d} = k_d / m$. Clearly, a signal with larger $\rho_d$ needs more measurements (larger $\delta_d$) to recover.

Let $\bD_d$ be a ${n_d} \times {n_1}$ downsampling matrix, $\bU_d$ be a ${n_1} \times {n_d}$ upsampling matrix, and ${\bx_d} = {\bD_d} \bx$ be the $n_d \times 1$ downsampled version of $\bx$. The LR-CS problem can be formulated as \cite{Bell}
\begin{equation}
\label{Model}
\begin{gathered}
  \by = \bA \bx + \bw
  = \bA({\bU_d}{\bx_d} + \bx - {\bU_d}{\bx_d}) + \bw \hfill \\
  {\text{  }} = {\bA \bU_d} {\bx_d} + {\bA}({\bI} - {\bU_d} {\bD_d})\bx + \bw, \hfill \\
\end{gathered}
\end{equation}
where $\bA \bU_d$ is the equivalent measurement matrix for the LR signal $\bx_d$, and ${\bA}({\bI} - {\bU_d}{\bD_d})\bx$ is the additional approximation error term when $\bx_d$ is the target signal to be recovered.  Note that this error term depends on the signal $\bx$.

The downsampling and upsampling matrices ${\bD_d}$ and ${\bU_d}$ play an important role in the MR-CS. In this paper, we require them to satisfy three conditions.

\begin{cond}
The downsampling and upsampling matrices ${\bD_d}$ and ${\bU_d}$ should be chosen such that if we first upsample a LR signal and then downsample to the original resolution, we can get back the original LR signal without any error. That is,
	\label{Cond1}
	\begin{equation}
	{\bD_d}{\bU_d} = {\bI_{n_d}}.
	\end{equation}
\end{cond}

Since $\bU_d$ is a tall matrix, this mild condition can be easily satisfied. In \cite{STONE}, the authors design a special two-resolution CS system such that a $m \times 1$ LR signal can be recovered directly from the $m \times 1$ CS sample $\by$. This can be considered as a special case of our setup.

The second condition is about the quality of the measurement matrix for the LR reconstruction.

\begin{cond}
\label{Cond_Qual}
The quality of the equivalent measurement matrix for the LR reconstruction should be no worse than that of the HR reconstruction.
\end{cond}

For different reconstruction algorithms, different criterions are used to evaluate the quality of the measurement matrix,  {\it e.g.}, the RIP constant for the basis pursuit algorithm \cite{RIP} and the mutual coherence for the orthogonal matching pursuit algorithm \cite{Coherence}. The solution in this paper is based on the AMP algorithm; hence we follow the requirement in  \cite{AMPoriginal,AMP,PTCPrediction,DAMP} that each entry of the LR measurement matrix should be i.i.d. Gaussian with zero mean and a variance of $1/m$.

Since $m < n_d$ in our case, the MR-CS problem here cannot be solved directly without exploiting the structure of ${\bx_d}$. Moreover, the LR signal should be easier to recover than the HR signal, {\it i.e.}, the amount of useful information ${k_d}$ contained in ${\bx_d}$ should be no more than the amount ${k_1}$ in the original HR signal $\bx$. We therefore also require the downsampling matrix ${\bD_d}$ to satisfy the following condition.

\begin{cond}
\label{Cond_Fam}
If $\bx$ belongs to the family ${\mathcal{F}_{n_1,\varepsilon_1}}$ in basis ${\bPsi}$, the downsampling matrix ${\bD_d}$ should be chosen such that ${\bx_d} = {\bD_d} \bx$ belongs to the family ${\mathcal{F}_{{n_d},{\varepsilon_d} }}$ in basis ${\bPsi_d} = \{ {\bD_d} {\bPsi} \}  - \{ \bf{0} \} $ with ${\varepsilon _d} \leqslant d{\varepsilon _1}$.
\end{cond}

Some results similar to Cond. \ref{Cond_Fam} have been reported in \cite{Bell} for simple sparse vectors, which is a special case of Cond. \ref{Cond_Fam}, as summarized below.

\begin{cond}
\label{Cond2Simp}
If $\bx$ is sparse in basis ${\bPsi}$, then ${\bx_d}={\bD_d} \bx$ is sparse in the non-zero projected low-dimension basis ${\bPsi_d} = \{ {\bD_d} {\bPsi} \}  - \{ \bf{0}\} $. The sparsity $k_d$ of ${\bx_d}$ is no larger than $k$, the sparsity of $\bx$, if the columns of ${\bPsi_d}$ are linearly independent.
\end{cond}
%

Our condition in Cond. \ref{Cond_Fam} is not restricted to simple sparse vectors, and can be used for other special structures that $\bx$ follows, such as piecewise constancy.

In Sec. \ref{MRImage}, we will design three pairs of up-/down-sampling matrices for images that satisfy the three conditions above perfectly or approximately. One pair is for simple sparse signals and two pairs are for piecewise constant signals. The conditions listed above can also be used to design matrices for the multi-resolution reconstructions of other types of structured sparse signals.

Note that the term "multi-resolution" in our paper is slightly different from that in the wavelet transform literature, because our method only reconstructs each of these LR signals independently, and how to use a LR reconstruction to help a HR reconstruction is not addressed in this paper. Nevertheless, we will show in Table \ref{L2H} that by simply upsampling the recovered LR image to the target HR, we can sometimes provide better HR image than reconstructing the HR image directly from the measurements.


\section{Multi-Resolution Approximate Message Passing}
\label{sec_MRAMP}

In this section, we propose an approximate message passing (AMP)-based algorithm to solve the MR-CS problem. Without loss of generality, we assume that the signal belongs to the structured sparse family ${\mathcal{F}_{n_1,\varepsilon_1}}$ in the canonical basis.

The main idea of the original AMP is to transform the CS reconstruction problem into a denoising problem \cite{PTCPrediction}, {\it i.e.}, estimating ${\bx_o}$ from its noisy observations ${\bx_o} + \sigma \be$, where entries of $\be$ are i.i.d. Gaussian with zero mean and unit variance, and $\sigma$ is a constant. In each iteration of AMP, a pseudo-data ${\bz^t} = {\bx^t} + {\bA^T} {r^t}$ is first formed. It is then denoised by a denoising function ${\eta _{{\sigma ^t}}}({{\bz}^t};\tau )$, where ${{\sigma ^t}}$ is the standard deviation (std) of ${\bz^t}$ and $\tau $ is the tuning parameter of the denoiser. Finally the residual of the measurements is updated. That is,
\begin{equation}
\label{EquAMPHR}
\begin{gathered}
  {\bz^t} = {\bx^t} + {\bA^T} {\br^t}, \hfill \\
  {\bx^{t + 1}} = {\eta _{{\sigma ^t}}}({\bz^t};\tau ), \hfill \\
  {\br^{t + 1}} = \by - {\bA}{\bx^{t + 1}} + {b^t}{\br^t}, \hfill \\
\end{gathered}
\end{equation}
where ${b^t}$ is the Onsager term, which is related to the divergence of the denoiser by
 \begin{equation}
\label{eq_onsager}
{b^t} =
\frac{1}{m}{\text{div}}\eta _{{\sigma^{t-1}}}(\bu;\tau){|_{\bu = {\bz_d^{t - 1}}}}
=
\frac{1}{m} \sum_{i=1}^{n_1} \frac{\partial \eta_{\sigma^{t - 1}}(\bu;\tau)}{\partial u[i]}
{|_{\bu = {\bz^{t - 1}}}}.
\end{equation}

For different structured signals, different denoisers $\eta_{\sigma^t}(\cdot)$ should be used. For example, for simple sparse signals, the well-known soft-thresholding should be used, whereas total variation (TV) denoiser is more appropriate for piecewise constant signals \cite{PTCPrediction}.

In order to apply AMP to the MR-CS problem in Eq. (\ref{Model}), we propose the following multi-resolution approximate message passing algorithm (MR-AMP),
\begin{equation}
\label{EquAMPLR}
\begin{gathered}
  \bz_d^t = \bx_d^t + {\bA_d^T} \br_d^t, \hfill \\
  \bx_d^{t + 1} = {\eta _{\sigma _d^t}}(\bz_d^t;\tau ), \hfill \\
  \br_d^{t + 1} = \by - {\bA_d} \bx_d^{t + 1} + b_d^t\br_d^t, \hfill \\
\end{gathered}
\end{equation}
where ${\bA}_d = {\bA \bU_d} \boldsymbol{\bf{\Lambda}} $ is the corresponding measurement matrix for the LR reconstruction, with $\bf{\Lambda} $ being a diagonal matrix determined by the upsampling matrix ${\bU_d}$ to normalize the columns of ${\bA \bU_d}$. $b_d^t$ is similar to Eq. (\ref{eq_onsager}) except that $n_1$ becomes $n_d$. Instead of estimating ${\bx_o}$, we are trying to estimate ${\bx_{d,o}} = {\bD_d} {\bx_o}$ from the pseudo-data $\bz_d^t$ with std $\sigma _d^t$ using the denoising function $\eta_{\sigma^t_d}(\cdot)$.

The original AMP in Eq. (\ref{EquAMPHR}) is a special case of MR-AMP in Eq. (\ref{EquAMPLR}) with $d=1$. In this paper, we denote the original AMP as high-resolution approximate message passing (HR-AMP), and MR-AMP with $d > 1$ as low-resolution approximate message passing (LR-AMP). Since the dimensions of ${\bA_d}$ and ${\bx_d}$ are smaller than those of ${\bA}$ and $\bx$, the complexity of LR-AMP is thus lower than HR-AMP.  Note that the proposed MR-AMP does not impose any additional constraint to the measuring matrix $\bA$ in the original AMP. It only modifies the reconstruction algorithm to get different LR estimates of the signal.

\section{State Evolution and Phase Transition of MR-AMP}
\label{PTC}

In this section, we analyze the theoretical performance of the proposed MR-AMP in terms of its state evolution, phase transition, and noise sensitivity.

\subsection{State Evolution}
\label{sec_se}

The availability of the state evolution analysis is an important advantage of AMP over many other CS algorithms. Empirical findings show that the MSEs of AMP with various denoisers can be predicted accurately by its state evolution \cite{DAMP,PTCPrediction}, which describes the asymptotic limit of the AMP estimates in Eq. (\ref{EquAMPHR}) when $m,{\text{ }}{n_1} \to \infty $, for any fixed $t$ \cite{AMPoriginal}. Starting from ${\theta ^0} = {\left\| {{\bx_o}} \right\|_2^2} / {n_1}$, the state evolution generates a sequence of numbers through the following iterations.
\begin{equation}
\label{SE_HR}
\begin{split}
{({\sigma ^t})^2} &= \frac{1}{{{\delta _1}}}{\theta ^t}({\bx_o},\delta_1 ,\sigma _w^2, \tau) + \sigma _w^2, \\
{\theta ^{t + 1}}({\bx_o},{\delta_1} ,\sigma _w^2,\tau) &= \frac{1}{n_1}\mathbb{E}\left\| {{\eta _{{\sigma ^t}}}({\bx_o} + {\sigma ^t}\be;\tau ) - {\bx_o}} \right\|_2^2,
\end{split}
\end{equation}
where the expectation is with respect to $\be \sim \mathcal{N}(0,{\bI})$. For large values of $m$ and $n_1$, the state evolution predicts the MSE of the AMP algorithm in Eq. (\ref{EquAMPHR}), {\it i.e.}, ${\theta ^t}({\bx_o},{\delta _1},\sigma _w^2,\tau) \approx \frac{1}{n_1}\left\| {{\bx^t} - {\bx_o}} \right\|_2^2$.

To get the state evolution of the proposed MR-AMP, we start from $\theta _d^0 = {{\left\| {{\bx_{d,o}}} \right\|_2^2}}/{{{n_d}}}$, where ${\bx_{d,o}}$ is the target LR signal. Let ${\sigma^2_{d,w}}$ denote the variance of the MR-AMP noise in Eq. (\ref{Model}), including contributions from the approximation error and measurement noise, which is equal to ($\sigma _w^2 + 1/m\left\| {({{\bI}} - {{{\bU}}_d}{{{\bD}}_d})\bx} \right\|_2^2$), as will be shown in Sec. \ref{sec_NS}. The state evolution of the MR-AMP is thus given by the following iterations.
\begin{equation}
\label{SE_LR}
\begin{split}
{(\sigma _d^t)^2} &= \frac{1}{{{\delta _d}}}\theta _d^t({\bx_{d,o}},{\delta _d},\sigma _{d,w}^2,\tau) + \sigma _{d,w}^2,\\
\theta _d^{t + 1}({\bx_{d,o}},{\delta _d},\sigma _{d,w}^2,\tau) &= \frac{1}{{{n_d}}}\mathbb{E}\left\| {\eta_{\sigma^t_d} ({\bx_{d,o}} + \sigma _d^t \be;\tau) - {\bx_{d,o}}} \right\|_2^2,
\end{split}
\end{equation}
 where $\sigma _d^t$ is the predicted std of the estimate $\bz_d^t$ in Eq. (\ref{EquAMPLR}). If $d=1$, Eq. (\ref{SE_LR}) reduces to that of AMP in Eq. (\ref{SE_HR}).

 Note that the state evolution of AMP is only proved rigorously for scalar denoisers, but not for non-scalar denoisers, such as total-variation-based denoisers and other more advanced denoisers \cite{DAMP,NCSUAMP,TvampParis}. However, similar to observations in these papers, empirical findings in Sec. \ref{sec_experiments} show that in all cases studied in this paper the MSEs of the MR-AMP can be predicted accurately by the state evolution above.

\subsection{Noiseless Phase Transition of LR-AMP}
\label{noiseless}

In CS reconstruction without sampling noise, the phase transition curve (PTC) defines the minimum number of CS measurements required to perfectly recover ${\bx_o}$, {\it i.e.}, ${\theta ^\infty }({\bx_o},{\delta _1},0,\tau) \to 0$ \cite{PTCPrediction}. In this part, we investigate the noiseless phase transition of MR-AMP, where we assume both $\sigma _w^2 = 0$ and $\left\| {{\bA}({\bI} - {\bU_d} {\bD_d})\bx} \right\|_2^2 = 0$ in Eq. (\ref{Model}). The latter is possible for some special signals, and an example will be given in Sec. \ref{sec_experiments}. We will show that by allowing LR reconstruction, the MR-AMP admits a family of PTCs, thereby enabling perfect reconstruction of a LR signal in the infeasible region of the original HR-AMP. This is an important generalization of the AMP theory.

The family ${\mathcal{F}_{n,\varepsilon }}$ is scale-invariant \cite{PTCPrediction}, {\it i.e.}, $\eta_{\sigma}(\by;\tau)=\sigma \eta_1(\by/\sigma;\tau)$. Therefore we only need to consider $\sigma = 1$, and we can simplify the notation ${\eta _{\sigma}}(\by;\tau)$ as $\eta (\by;\tau)$. We then define the following asymptotic minimax MSE when a denoiser $\eta $ with parameter $\tau$ is used to recover signals in the structured sparse family ${\mathcal{F}_{{n_1},{\varepsilon_1} }}$ \cite{PTCPrediction}.
\begin{equation}
\label{MinimaxMSE}
%
M({\varepsilon _1}|\eta ) \equiv
\mathop {\lim }\limits_{{n_1} \to \infty } \frac{1}{{{n_1}}}\mathop {\inf }\limits_\tau  \mathop {\sup }\limits_{{v_{{n_1}}} \in {{\cal F}_{{n_1},{\varepsilon _1}}}} {\mathbb{E}_{{v_{{n_1}}}}}\left\| {{\eta }({{\bf{x}}_o} + {\bf{e}};\tau ) - {\rm{ }}{{\bf{x}}_o}} \right\|_2^2,
\end{equation}
In words, $M({\varepsilon _1}|\eta )$ is obtained by tuning the denoiser parameter to minimize the MSE per coordinate of the least favorable distribution in the family. The tuning rules of the parameters $\tau $ are provided in Sec. \ref{TuningRule}.


{ The minimax MSE has some basic properties \cite{AMP,PTCPrediction}. First, since the denoising can improve the reconstruction, we have $0 \le M({\varepsilon _1}|\eta ) \le 1$. Besides, $M({\varepsilon _1}|\eta ) \to 0$ when $\varepsilon _1 \to 0$, and $M({\varepsilon _1}|\eta ) \to 1$ when $\varepsilon _1 \to 1$. Second, $M({\varepsilon _1}|\eta )$ is monotonically increasing with respect to $\varepsilon_1$ \cite{PTCPrediction}, because the reconstruction difficulty increases with $\varepsilon_1$.}

The detailed expression of $M(\varepsilon_1 |\eta )$ for AMP with various denoisers is derived in \cite{AMP,PTCPrediction,AMPoriginal}. More importantly, it is shown in \cite{PTCPrediction} that $M(\varepsilon_1 |\eta )$ defines the minimum CS undersampling ratio $\delta_1$ for perfect reconstruction, {\it i.e.}, it describes the phase transition curve of AMP as follows.

\begin{theorem}
\label{PTCHR}
In the noiseless case, when using AMP with denoiser $\eta$ to reconstruct signals in ${\mathcal{F}_{{n_1},{\varepsilon_1} }}$, the AMP succeeds with high probability if
\begin{equation}
\label{PTCHREqu}
{\delta_1} > M({\varepsilon_1}|\eta ).
\end{equation}
Viceversa AMP fails with high probability for ${\delta_1}  < M({\varepsilon_1} |\eta )$.
\end{theorem}

Combining Theorem \ref{PTCHR} and the conditions in Sec. \ref{sec_bkgd}, we obtain the following generalized phase transition result for MR-AMP, which specifies the minimum sampling ratio to perfectly recover a LR signal. When $d=1$, it reduces to Theorem \ref{PTCHR}.

\begin{corollary}
\label{PTCLR}
{ When Cond. \ref{Cond1}, \ref{Cond_Qual} and \ref{Cond_Fam} are satisfied}, if a signal $\bx \in {\mathcal{F}_{{n_1},{\varepsilon_1} }}$ is sampled according to Eq. (\ref{CSeq}), and if $\sigma_w^2=0$ and $\left\| {({\bI} - {{\bU_d}}{{\bD_d}})\bx} \right\|_2^2=0$ in Eq. (\ref{Model}), then a LR signal $\bx_d \in {\mathcal{F}_{{n_d},{\varepsilon_d} }}$ with ${\varepsilon _d} \leqslant d \, {\varepsilon _1}$ can be reconstructed perfectly with high probability via the LR-AMP in Eq. (\ref{EquAMPLR}) when the CS undersampling ratio satisfies
\begin{equation}
\label{PTCLREqu}
{\delta_1}  > M(d{\varepsilon_1} |\eta )/d,
\end{equation}
where $M(\varepsilon_1|\eta)$ is the minimax MSE of the original HR-AMP. Viceversa the LR-AMP fails with high probability for ${\delta_1}  < M(d{\varepsilon_1} |\eta )/d$.
\end{corollary}

\begin{IEEEproof}
As mentioned before, ${\delta_d} = d{\delta_1}$. Since there is no approximation error in Eq. (\ref{Model}), Theorem \ref{PTCHR} can be applied directly to the LR-AMP. Therefore the LR-AMP succeeds with high probability if the CS sampling ratio satisfies
\begin{equation*}
{\delta _d} = d{\delta_1}  > M({\varepsilon _d}|\eta ).
\end{equation*}

If Cond. \ref{Cond_Fam} is satisfied, we have ${\varepsilon _d} \leqslant d{\varepsilon_1} $. Eq. (\ref{PTCLREqu}) can thus be obtained using the property that $M({\varepsilon _d}|\eta )$ is monotonically increasing with respect to ${\varepsilon _d}$.
\end{IEEEproof}


{ The next result shows that the LR reconstruction requires less sampling rate than the HR-AMP. That is, the LR-AMP has larger feasible operating region than the original HR-AMP under certain condition.

\begin{corollary}
\label{Concave}
If $M({\varepsilon _1}|\eta )$ is a concave function of ${\varepsilon _1}$, then we have $M(d{\varepsilon _1}|\eta )/d \le M({\varepsilon _1}|\eta )$.
\end{corollary}

\begin{IEEEproof}
It is known that if a function $f$ is concave, and $f(0) \ge 0$, then $f$ is subadditive, {\it i.e.}, $f(x+y) \le f(x) + f(y)$. From this we can get $ f(tx) \le tf(x)$ for $t \ge 1$. It is clear from the definition that $M(\varepsilon_1|\eta) \ge 0$. Therefore if $M(\varepsilon_1|\eta)$ is concave, then by the subadditivity, we can get $M(d{\varepsilon _1}|\eta ) \le dM({\varepsilon _1}|\eta )$, {\it i.e.}, $M(d{\varepsilon _1}|\eta )/d \le M({\varepsilon _1}|\eta )$.
\end{IEEEproof}

The concavity condition of $M({\varepsilon _1}|\eta )$ is satisfied for many families of structured signals. In particular, it is proved in \cite{Minimax94} for simple sparse signals in Eq. (\ref{SimpleSparse}) when the soft-thresholding denoiser is used. It is also confirmed in \cite{PTCPrediction} for block-sparse signals with block soft-thresholding denoiser. In the Appendix, we prove that it is satisfied for piecewise constant signals. Finally, we also show in Sec. \ref{MRImage} that the concavity condition holds for 2D images in both the simple sparse and piecewise constant families.

}

Corollary \ref{Concave} confirms the motivation discussed in the introduction of the paper, {\it i.e.}, if the CS sampling rate is too low, although the full-resolution reconstruction will fail, we can still reconstruct a LR version of the signal. Moreover, in the noiseless case, given ${\delta _1}$, ${\varepsilon _1}$, we can precisely determine the critical downsampling factor $d$ by solving the equation ${\delta _1} = M(d{\varepsilon _1}|\eta )/d$.



\subsection{Noise Sensitivity of MR-AMP}
\label{sec_NS}

{ The noiseless case studied above is quite restrictive. In practices, we are more interested in the performance of the algorithm in the presence of noise.} In this part, we study the noise sensitivity of LR-AMP when the noises $\bw$ and ${{\bA}({\bI} - {\bU_d} {\bD_d})\bx}$ in Eq. (\ref{Model}) are not zero. As in  \cite{AMP,DAMP}, the noise sensitivity of HR-AMP is defined as
\begin{equation*}
NS(\sigma _w^2,{\delta _1}) = \mathop {\inf }\limits_\tau  \mathop {\sup }\limits_{{v_{{n_1}}} \in {F_{{n_1},{\varepsilon _1}}}} {\mathbb{E}_{{v_{{n_1}}}}}\{ {\theta ^\infty }({{\bf{x}}_o},{\delta _1},\sigma _w^2,\tau )\} ,
\end{equation*}
which is the minimax MSE per coordinate of the HR-AMP output when the iteration number goes to $\infty$ in Eq. (\ref{SE_HR}). {  It is shown in \cite{AMP,DAMP} that when the undersampling ratio meets the same phase transition condition as in Theorem \ref{PTCHR}, the structured sparse signal can be recovered with a bounded noise sensitivity.}

When studying the noise sensitivity of the LR-AMP, we use $NS(\sigma _{d,w}^2,{\delta _d})$ to represent the noise sensitivity of LR-AMP, where $\sigma _{d,w}^2$ is the variance of LR-AMP noise. { The next result shows that when the undersampling ratio meets the same condition as in Corollary \ref{PTCLR}, we can also recover the LR signal $\bx_d$ with a bounded noise sensitivity.}

\begin{corollary}
\label{NSLR}
{ When Cond. \ref{Cond1}, \ref{Cond_Qual} and \ref{Cond_Fam} are satisfied}, if the undersampling ratio satisfies Eq. (\ref{PTCLREqu}), {\it i.e.}, ${\delta _1}  > M(d{\varepsilon _1} |\eta )/d$ in the compressed sensing of $\bx \in {\mathcal{F}_{{n_1},{\varepsilon_1} }}$  in Eq. (\ref{CSeq}) with noise variance $\sigma_w^2$, a LR version of the signal $\bx_d \in {\mathcal{F}_{{n_d},{\varepsilon_d} }}$ with ${\varepsilon _d} \leqslant d{\varepsilon _1}$ can be reconstructed via LR-AMP with downsampling matrix ${\bD}_d$ and upsampling matrix ${\bU_d}$, and the noise sensitivity is bounded by
\begin{equation}
\label{EquNSLR}
\begin{split}
&NS(\sigma _{d,w}^2,{\delta _d}) \\
&\leqslant \frac{{M(d{\varepsilon _1} |\eta )}}{{1 - M(d{\varepsilon _1} |\eta )/(d{\delta _1} )}}(\sigma _w^2 + \frac{1}{m}\left\| {({\bI} - {\bU_d} {\bD_d}) \bx} \right\|_2^2).
\end{split}
\end{equation}
\end{corollary}

\begin{IEEEproof}
According to Prop. 2 in \cite{DAMP}, the noise sensitivity of AMP with various denoisers is bounded by
\begin{equation*}
NS(\sigma _w^2,{\delta_1} ) \leqslant \frac{{M({\varepsilon_1} |\eta )}}{{1 - M({\varepsilon_1} |\eta )/{\delta_1} }}\sigma _w^2.
\end{equation*}

Replacing ${\delta _1}$, ${\varepsilon _1}$ and $\sigma _w^2$ by ${\delta _d}$, ${\varepsilon _d}$ and $\sigma _{d,w}^2$ in the formula above respectively, we have

\begin{equation*}
NS(\sigma _{d,w}^2,{\delta _d}) \le \frac{{M({\varepsilon _d}|\eta )}}{{1 - M({\varepsilon _d}|\eta )/{\delta _d}}}(\sigma _w^2 + \sigma _{d,w}^2).
\end{equation*}

Since ${M({\varepsilon _d}|\eta )}$ is monotonically increasing with ${{\varepsilon _d}}$, it is easy to see that $\frac{{M({\varepsilon _d}|\eta )}}{{1 - M({\varepsilon _d}|\eta )/{\delta _d}}}$ is also monotonically increasing. Together with $ \varepsilon_d \le d \varepsilon_1$, we can have

\begin{equation*}
NS(\sigma _{d,w}^2,{\delta _d}) \le \frac{{M(d{\varepsilon _1}|\eta )}}{{1 - M(d{\varepsilon _1}|\eta )/(d{\delta _1})}}(\sigma _w^2 + \sigma _{d,w}^2).
\end{equation*}

By the central limit theorem, if the entries of ${\bA}$ have i.i.d. $\mathcal{N}(0, 1/m)$ distribution, and ${\bD_d}$ and ${\bU_d}$ are deterministic, then for a given $\bx$, each entry of ${\bA}({\bI} - {\bU_d} {\bD_d})\bx$ converges to i.i.d. Gaussian distribution with zero mean and variance $1/m\left\| {({\bI} - {\bU_d} {\bD_d})\bx} \right\|_2^2$. Therefore the equivalent noise variance $\sigma _{d,w}^2$ for the LR-AMP problem is ($\sigma _w^2 + 1/m\left\| {({\bI} - {\bU_d} {\bD_d})\bx} \right\|_2^2$), which proves the result.
\end{IEEEproof}

{ Different from the original AMP, the upper bound of the LR-AMP noise sensitivity $NS(\sigma _{d,w}^2,{\delta _d})$  is conditional, since it depends on the approximation error term $({\bI} - {{\bU_d}{\bD_d}})\bx$, which varies for different input signals. Therefore it is crucial to design good up-/down-sampling matrices to reduce the LR reconstruction error, which will be studied in Sec. \ref{MRImage}. It should be noted that the upper bound is finite in many applications. Moreover, sometimes we can further derive a signal-independent upper bound. For example, in 8-bit images, the pixel value ranges from $0$ to $255$. Therefore the worst value of each entry in $({\bI} - {{\bU_d}{\bD_d}})\bx$ is $255$, and the worst value of $\left\| {({\bI} - {{\bU_d}{\bD_d}})\bx} \right\|_2^2$ is thus ${255^2}{n_1}$. The upper bound in Eq. (\ref{EquNSLR}) can be further bounded by
\begin{equation}
\begin{split}
&NS(\sigma _{d,w}^2,{\delta _d}) \le \frac{{M(d{\varepsilon _1}|\eta )}}{{1 - M(d{\varepsilon _1}|\eta )/(d{\delta _1})}}(\sigma _w^2 + \frac{{{255^2}}}{{{\delta _1}}})\\
& \qquad \qquad \quad \le \frac{{M(d{\varepsilon _1}|\eta )}}{{1 - M(d{\varepsilon _1}|\eta )/(d{\delta _1})}}(\sigma _w^2 + \frac{{{255^2}d}}{{M(d{\varepsilon _1}|\eta )}}).
\end{split}
\end{equation}
The upper bound above is too pessimistic since the LR approximation ${{\bU_d}{\bD_d}}\bx$ usually has much less approximation error than $255$. The upper bound can be reduced if more accurate estimate of  $\left\| {({\bI} - {{\bU_d}{\bD_d}})\bx} \right\|_2^2$ is known.
}

Corollary \ref{NSLR} is more general than Corollary \ref{PTCLR}, as it allows sampling noise and LR approximation noise. It gives further affirmative answers to the questions raised in the introduction of the paper, {\it i.e.}, if the CS sampling rate is too low for the full-resolution signal recovery, we can reconstruct a LR version of the signal with bounded noise sensitivity. The noisy case shares the same PTC with the noiseless one, as in the original AMP, which serves as a guideline to determine the critical resolution under which the noise sensitivity of the LR signal recovery is bounded.

\section{Design of Downsampling and Upsampling Matrices for MR-AMP}
\label{MRImage}

In this section, we give examples on the design of the up-/down-sampling matrices that satisfy the three conditions in Sec. \ref{sec_bkgd} perfectly or approximately, so that they can be used in MR-AMP-based image reconstruction. Three pairs of matrices will be designed. The first pair is in the DCT or wavelet transform domain and is designed for the simple sparse family. The other two pairs are in the spatial domain and are suitable for piecewise constant signals.

In \cite{Bell}, DCT-based and total-variation (TV)-based up-/down-sampling matrices are designed for videos such that the downsampling matrix ${\bD_d}$ satisfies Cond. \ref{Cond2Simp} and the upsampling matrix ${\bU_d}$ satisfies Cond. \ref{Cond1}. However, the proof in it is mainly about TV-based up-/down-sampling matrices. Moreover, the impact of MR design on the quality of the measurement matrix is not considered, {\it i.e.}, it is not clear whether Cond. \ref{Cond_Qual} holds or not.

\subsection{Transform-Domain Downsampling and Upsampling}
\label{sec_trunc}

Natural images are approximately sparse in DCT or wavelet domain. The sparse representation of a $n_1 \times n_1$ image ${\bX}$ thus belongs to the simple sparse family in Eq. (\ref{SimpleSparse}), and the soft-thresholding denoiser can be used in the transform domain. To apply CS sampling and reconstruction to images, we need to introduce the transform basis to Eq. (\ref{CSeq}) and Eq. (\ref{Model}).

For a ${n_1} \times {n_1}$ image ${\bX}$, a ${n_d} \times {n_d}$ LR image ${\bX_d}$ can be obtained via transform-domain downsampling by first applying HR 2D transform, extracting the $n_d \times n_d$ low-frequency coefficients, and then applying the LR 2D inverse transform \cite{Trac07,ImgResize}.

Let ${\bPsi_{n_1}}$ and ${\bPsi_{n_d}}$ represent the $n_1 \times n_1$ and $n_d \times n_d$ DCT or orthogonal multiple-level wavelet transform respectively. We use the following 1D transform-domain downsampling operator \cite{Trac07}
\begin{equation}
\label{DdEq}
{\bD_d} = \sqrt{\frac{1}{d}} \, \bPsi_{n_d}^T {\bI_{n_d \times n_1}} {\bPsi_{n_1}}.
\end{equation}
where the fat identity matrix ${\bI_{n_d \times n_1}}$ serves as a truncation operator, because it only keeps the first $n_d$ coefficients of the input after being transformed by ${\bPsi_{n_1}}$.

Given the downsampling matrix, one way to satisfy Cond. \ref{Cond1}, {\it i.e.,} ${\bD_d} {\bU_d} = {\bf{I}}$, is to use transform-domain zero-padding. The corresponding upsampling matrix ${\bU_d}$ is
\begin{equation}
{\bU_d} =   \sqrt{d} \, \bPsi_{n_1}^T {\bI_{n_1 \times n_d}} {\bPsi_{n_d}}.
\end{equation}

The 2D downsampling and upsampling can thus be represented as
\begin{equation}
\label{EqUpDown2D}
\begin{split}
{\bX_d} &= {\bD_d} {\bX} \bD_d^T,\\
{{\hat \bX}} &= {\bU_d} {\bX_d} \bU_d^T.
\end{split}
\end{equation}

It should be noted that according to the definitions in \cite{ImgResize}, for the downsampling in DCT domain, we can achieve non-integer downsampling ratio since we simply take the top left ${n_d} \times {n_d}$ low-frequency coefficients and apply the LR 2D inverse DCT transform. However, for the downsampling in wavelet domain, we can only get integer downsampling ratio that is power of 2, since the LR image is the appropriately scaled low-pass subband in the multi-level wavelet transform.

Let $\bx$, $\bx_d$, and $\hat{\bx}$ be the vectorized versions of ${\bX}$, ${\bX_d}$, ${{\hat \bX}}$, respectively, by concatenating the columns of each matrix together. Let $\otimes$ denote the Kronecker product, the 2D downsampling and upsampling can be converted to the following 1D formulas.
\begin{equation}
\label{KronUpDown}
\begin{split}
\bx_d &= ({\bD_d} \otimes {\bD_d}) \bx, \\
\hat{\bx} &= ({\bU_d} \otimes {\bU_d}) \bx_d.
\end{split}
\end{equation}

Similarly, let ${\bS_1} =  {\bPsi_{n_1}} {\bX} {\bPsi_{n_1}^T}$ and ${\bS_d} = {\bI_{n_d \times n_1}} {\bS_1}  {\bI_{n_1 \times n_d}}$ be the 2D transform of ${\bX}$ and its low-frequency part, and $\bs_1$ and $\bs_d$ be their vectorized versions. The 2D inverse transform can be represented by 1D transform as follows
\begin{equation}
\begin{gathered}
  \bx = ({\bPsi_{n_1}^T} \otimes {\bPsi_{n_1}^T}){\bs _1}, \hfill \\
  {\bx_d} = \frac{1}{d}({\bPsi_{n_d}^T} \otimes {\bPsi_{n_d}^T}){\bs _d}, \hfill \\
\end{gathered}
\end{equation}
where the two matrices are still orthogonal. Note that the corresponding 1D downsampling ratio is $n_1^2/n_d^2=d^2$. {It is easy to see that the concavity condition in Corollary \ref{Concave} holds here, since the 1D sparse representation of a 2D image is just the vectorized version of its 2D representation.}

We next show that the transform-domain up-/down-sampling operators defined above satisfy Cond. \ref{Cond_Qual} and Cond. \ref{Cond_Fam}.

First, we assume ${\bs _1} \in \mathcal{F}_{n_1^2,{\varepsilon _1}}^{SS}$. Since the transform-domain downsampling operator simply extracts the low-frequency components of ${\bs _1}$, the number of nonzero entries in ${\bs _d}$  is certainly no more than that in ${\bs _1}$; hence $\varepsilon_d \le n_1^2 \varepsilon_1 / n_d^2 = d^2 \varepsilon_1$, and ${\bs _d} \in \mathcal{F}_{n_d^2, d^2 {\varepsilon _1}}^{SS}$. Cond. \ref{Cond_Fam} is thus satisfied.

To check Cond. \ref{Cond_Qual}, note that the equivalent 1D measurement matrix for the HR signal is ${\bPhi_1} = {\bA}({\bPsi_{n_1}^T} \otimes {\bPsi_{n_1}^T})$, whereas the equivalent 1D measurement matrix for the LR-CS problem in Eq. (\ref{Model}) is
\begin{equation}
\label{ScaleTrans}
\begin{split}
  {\bPhi_d} &= \frac{1}{d}{\bA}({\bU_d} \otimes {\bU_d})({\bPsi_{n_d}^T} \otimes {\bPsi_{n_d}^T}) \\
   &= {\bA}({\bPsi_{n_1}^T} {\bI_{{n_1} \times {n_d}}}) \otimes ({\bPsi_{n_1}^T} {\bI_{{n_1} \times {n_d}}}).
\end{split}
\end{equation}
It is easy to see that ${\bPhi_d}$ is the first ${n_d^2}$ columns of ${\bPhi_1}$. Since our proposed algorithms are based on AMP, where each entry of the measurement matrix ${\bA}$ follows i.i.d. $\mathcal{N}(0, 1/m)$ distribution, it can be shown that given ${\bPsi_{n_1}}$, each entry of ${\bPhi_1}$ and ${\bPhi_d}$ also has i.i.d. $\mathcal{N}(0, 1/m)$ distribution. Therefore, with the proposed transform-domain up-/down-sampling method, the quality of the measurement matrix for the LR-AMP is the same as that of the HR-AMP.

\subsection{Spatial-Domain Downsampling and Upsampling}
\label{sec_sum}

We next develop two pairs of spatial-domain up-/down-sampling matrices for MR-AMP. In this part, we assume images are piecewise constant and belong to the family ${\mathcal{F}^{PC}_{{n_1},{\varepsilon_1}}}$ in Eq. (\ref{Piecewise}), which has a small number of change points.

\subsubsection{Solution 1}

We first design the operators for 1D signals and then extend them to 2D images. For 1D piecewise constant signals, to satisfy Cond. \ref{Cond_Fam}, the first downsampling matrix ${\bD_d}$ we use is the row-decimated identity matrix, {\it i.e.}, a matrix whose $(i,di)$-th entries are $1$ for all $i$, and all other entries are zero. The downsampled signal can be written as
\begin{equation}
\label{TVxd}
\bx_d = {\bD_d} \bx
= {\left[ {\begin{array}{cccc}
  {{x[d]}}&{{x[2d]}}&{\hdots}&{{x[{n_d}d]}}
\end{array}} \right]^T},
\end{equation}
where $x[i]$ represents the $i$-th entry of $\bx$.

The corresponding upsampling matrix ${\bU_d}$ used in this part is the { repetition} operator which duplicates each input sample by $d$ times.
\begin{equation}
\label{EquRepetition}
{\bU_d} =
\left[ {\begin{array}{*{20}{c}}
  {{\bf{1}}_{d \times 1}}&& \\
  &\ddots& \\
  &&{{\bf{1}}_{d \times 1}}
\end{array}} \right],
\end{equation}
where ${{\bf{1}}_{d \times 1}}$ is an all-one vector. Clearly ${\bD_d}$ and ${\bU_d}$ satisfy ${\bD_d} {\bU_d} = {\bI}$ in Cond. \ref{Cond1}.



Next, we show that the spatial-domain up-/down-sampling matrices also satisfy Cond. \ref{Cond_Fam}.

\begin{lemma}
\label{Cond2TV}
If $\bx$ is a piecewise constant signal generated from the family ${\mathcal{F}^{PC}_{{n_1},{\varepsilon_1}}}$ in Eq. (\ref{Piecewise}), then the downsampled signal ${\bx_d}$ in Eq. (\ref{TVxd}) belongs to ${\mathcal{F}^{PC}_{{n_d},{\varepsilon _d}}}$ with ${\varepsilon _d} \leqslant d{\varepsilon _1}$.
\end{lemma}

\begin{IEEEproof}

A $n_1 \times 1$ piecewise constant signal $\bx$ is sparse in the differential domain.
\begin{equation}
\label{s1}
\small{
  {\bs _1} = \left[ {\begin{array}{*{20}{c}}
{ - 1}&1&{}&{}\\
{}&\ddots&\ddots&{}\\
{}&{}&{ - 1}&1
\end{array}} \right]\bx \equiv {{{\bPsi_{n_1}}}} \bx 
= \left[ {\begin{array}{c}
  {x[2] - x[1]} \\
   {x[3] - x[2]}\\
   \vdots  \\
  {x[{n_1}] - x[{n_1} - 1]}
\end{array}} \right]. \hfill \\
}
\end{equation}

Similarly, the representation of the downsampling signal in the differential domain can be written as
\begin{equation}
\label{sD}
\begin{gathered}
  {\bs _d} = {\bPsi _{n_d}}{\bx_d} \hfill \\
   = {\left[ {\begin{array}{*{20}{c}}
  {x[2d]-x[d],}&{\hdots,}&{x[{n_d}d] - x[({n_d} - 1) d]}
\end{array}} \right]^T}. \hfill \\
\end{gathered}
\end{equation}
Therefore, calculating the number of change points in $\bx_d$ is equivalent to counting the number of nonzero entries in $\bs_d$.

To facilitate the proof, we construct two new vectors $\bs_1^* = {\left[ {\begin{array}{*{20}{c}}
{x[1]}&{\bs_1^T}
\end{array}} \right]^T}$ and $\bs_d^* = {\left[ {\begin{array}{*{20}{c}}
{x[d]}&{\bs_d^T}
\end{array}} \right]^T}$, {\it i.e.}, adding the first entry of $\bx$ and $\bx_d$ to $\bs_1$ and $\bs_d$ respectively. {If we add $d$ consecutive entries of $\bs_1^*$, we can get one entry of $\bs_d^*$. For example, $(x[d + 1] - x[d]) + (x[d + 2] - x[d + 1]) + ... + (x[2d] - x[2d - 1]) = x[2d] - x[d]$. In matrix form, this means ${\bs _d^*} = \bU_d^T{\bs _1^*}$. If $\bx$ is generated from ${\mathcal{F}^{PC}_{{n_1},{\varepsilon_1}}}$, the maximum expected number of nonzero entries in $\bs_1^*$ will be ${n_1}{\varepsilon _1}+1$, due to the extra $x[1]$ in it. According to ${\bs _d^*} = \bU_d^T{\bs _1^*}$, the maximum expected number of nonzero entries in $\bs_d^*$ is still ${n_1}{\varepsilon _1}+1$. This happens when there is at most one nonzero entry in every $d$ entries in ${\bs _1^*}$; hence ${\varepsilon _d} \le ({n_1}{\varepsilon _1} + 1)/{n_d}$ $= d{\varepsilon _1} + 1/{n_d}$ $\to d{\varepsilon _1}$ when ${n_1} \to \infty $.}

\end{IEEEproof}

We next extend the results above to 2D images. The 2D ${n_d} \times {n_d}$ LR image ${\bX_d}$ can be written as Eq. (\ref{EqUpDown2D}) with ${\bD_d}$ in Eq. (\ref{TVxd}). If an image is piecewise constant, its 2D gradient is sparse, where the 2D gradient at each pixel is given by
\begin{equation}
{(\nabla {\bX})_{i,j}} = [{X_{i + 1,j}} - {X_{i,j}},{X_{i,j + 1}} - {X_{i,j}}].
\end{equation}

{The number of change points in a 2D piecewise constant signal $\bX$ equals to the number of nonzero entries in $\nabla \bX$, where ${(\nabla \bX)_{i,j}}$ is counted as one nonzero entry if one or two of its components are nonzero. Therefore, we can also vectorize the 2D $\nabla \bX$ into a 1D vector, and apply the method in the Appendix to prove the concavity in Corollary \ref{Concave} for 2D piecewise constant signals.} Additionally, the vertical differences and the horizontal differences are disjoint. By Lemma \ref{Cond2TV}, the number of horizontal or vertical change points of ${\bX_d}$ is no larger than that of ${\bX}$, thus Cond. \ref{Cond_Fam} is true for 2D images.

The remaining problem is to choose the appropriate denoiser for 2D piecewise constant signals. In this paper, instead of using the denoisers discussed in \cite{DAMP,NCSUAMP}, such as NLM (non-local means) and BM3D (3D block matching), we use a 2D-TV-based denoiser in ${\eta _{\sigma _d^t}}(\bz_d^t)$ of Eq. (\ref{EquAMPLR}). Our method is denoted as AMP-TV-2D.

The TV norm of 2D piecewise constant signals is defined as
\begin{equation}
\label{TV}
{\left\| {{\bX}} \right\|_{{\text{TV}}}} = \sum\limits_{i,j} {\sqrt {{{\left| {{X_{i + 1,j}} - {X_{i,j}}} \right|}^2} + {{\left| {{X_{i,j + 1}} - {X_{i,j}}} \right|}^2}} },
\end{equation}
which is isotropic and un-differentiable. { This norm will be used by the 2D-TV-based denoiser. Further details are given in Sec. \ref{TuningRule}}. It is different from the 1D TV denoiser in \cite{PTCPrediction} where TV norm for 1D piecewise constant signal is written as
${\left\| \bx \right\|_{\text{TV}}} = \sum\limits_{i = 1}^{{n_1} - 1} {\left| {{x_{i + 1}} - {x_i}} \right|} $.

{ In Sec. \ref{AMPandTVAL3}, we compare the performance of our AMP-TV-2D with the state-of-the-art algorithm TVAL3 (TV minimization by Augmented Lagrangian and ALternating direction ALgorithms) in \cite{TVAL3}. Note that TVAL3 depends on two slack parameters, which have to be manually tuned for each image and each measurement rate. In contrast, the thresholding parameters in our AMP-TV-2D are automatically tuned in each iteration, which will be discussed in Sec. \ref{TuningRule}. Recently, a similar algorithm to our AMP-TV-2D, the dual-constraints AMP (DC-AMP) (Sec. 8.1 of \cite{TvampParis}), is developed for 2D piecewise smooth signals, which can achieve similar performance to TVAL3. However, it also has a smoothness parameter that needs to be manually tuned. Moreover, there is no theoretical analysis for DC-AMP.}

Given the spatial-domain up-/down-sampling matrices, to satisfy Cond. \ref{Cond_Qual}, {\it i.e.}, the quality of the measurement matrix for LR-AMP is no worse than that of HR-AMP, we need to normalize the measurement matrix for LR-AMP, {\it i.e.},
\begin{equation}
\label{ScaleSpat}
{{{\bPhi_d}}} = \frac{1}{{d}}{{\bA}}({{{\bU_d}}} \otimes {{\bU_d}}),
\end{equation}
such that each entry of ${\bPhi_d}$ has i.i.d. $\mathcal{N}(0, 1/m)$ distribution.

\subsubsection{Solution 2}

In addition to the simple up-/down-sampling matrices in Eq. (\ref{TVxd}) and Eq. (\ref{EquRepetition}), we also develop a pair of bicubic up/-downsampling matrices and evaluate them in Sec. \ref{sec_2DExample}. In bicubic downsampling, each pixel in the LR image is the weighted average of sixteen pixels in the HR image, which has been known to produce smoother LR image than Eq. (\ref{TVxd}), {\it i.e.}, with less number of change points in ${\bX_d}$. Therefore Cond. \ref{Cond_Fam} holds for bicubic downsampling. On the other hand, the upsampling first inserts $d-1$ zeros between neighboring samples of the LR image and then performs bicubic interpolation. However, it can be verified that the corresponding product ${{\bD_d}} {{\bU_d}}$ is not an identity matrix, although very close. Therefore, strictly speaking, Cond. \ref{Cond_Qual} does not hold for bicubic matrices, and the simple scaling matrix $\bf{\Lambda}$ cannot make each entry of ${\bPhi_d}$ exactly having i.i.d. $\mathcal{N}(0,1/m)$ distribution. Nevertheless, this is still approximately true, and the efficiency of this scheme will be verified empirically in Sec. \ref{sec_2DExample}. Moreover, according to Corollary \ref{NSLR}, the conditional upper bound of the noise sensitivity is proportional to the LR approximation error $\left\| {({\bI} - {\bU_d} {\bD_d})\bx} \right\|_2^2$. Therefore for images, in terms of LR approximation error, the bicubic up-/down-sampling matrices are still better than the simple matrices in Eq. (\ref{TVxd}) and Eq. (\ref{EquRepetition}).

Finally, we point out the differences of our methods with those in \cite{Bell,ICIP14}. In \cite{Bell}, a similar spatial-domain up-/down-sampling framework was proposed, but the proof in it was implicit. Also, TVAL3 was chosen as the reconstruction algorithm, which requires manual tuning of two parameters. Moreover, the reconstruction performance cannot be predicted. Our AMP-TV-2D does not have manually tuned parameter, and its performance can be accurately predicted via state evolution. In \cite{ICIP14}, the same piecewise constancy model and the up-/down-sampling matrices as in Eq. (\ref{TVxd}) and (\ref{EquRepetition}) are used. It first reconstructs the original HR image and uses this estimated HR image to reduce the approximation error ${\bA} ({\bI} - {\bU_d} {\bD_d})\bx$. However, there is no theoretical guarantee that such operation can reduce the approximation error, and the algorithm only works when the undersampling rate ${\delta _1}$ is sufficiently large, at least $20\% $. Moreover, the complexity of this approach is higher than reconstructing the LR image directly.

\section{Experimental Results}
\label{sec_experiments}

In this section, we demonstrate the performance of the proposed MR-AMP with both transform-/spatial-domain up-/down-sampling, denoted by AMP-ST (soft thresholding) and AMP-TV (total variation), respectively. Empirical results will also be shown to verify some theoretical results. In each method, to facilitate comparison with the conventional approach, we use LR-AMP-ST and LR-AMP-TV to denote the proposed LR reconstruction schemes, and HR-AMP-ST and HR-AMP-TV to denote the original AMP with HR reconstruction. In addition, H2L-AMP-ST and H2L-AMP-TV represent the naive solutions that first reconstruct the HR signal and then downsample to the LR.

All tests in this paper use  column-normalized i.i.d. Gaussian measurement matrix ${\bA}$. All simulations are conducted on a PC with 3.4GHz Intel Core i7 quad-core processor and 64GB of memory. The testing images used include popular images Lena, Barbara, Boat, House, and Peppers, as well as some land remote sensing images, including the Memorial Stadium at the University of Nebraska Cornhuskers, and Sea World in San Diego. We follow the setup in \cite{DAMP} to rescale all images to $128 \times 128$. This enables the entire measurement matrix ${\bA}$ to be stored in the memory. We also include some experiments of larger $256 \times 256$ images to demonstrate the visual comparison, following the same setup in \cite{DAMP}.

\begin{figure}[tb]
\begin{center}
\begin{tabular}{c}
  \includegraphics[width=2.7in]{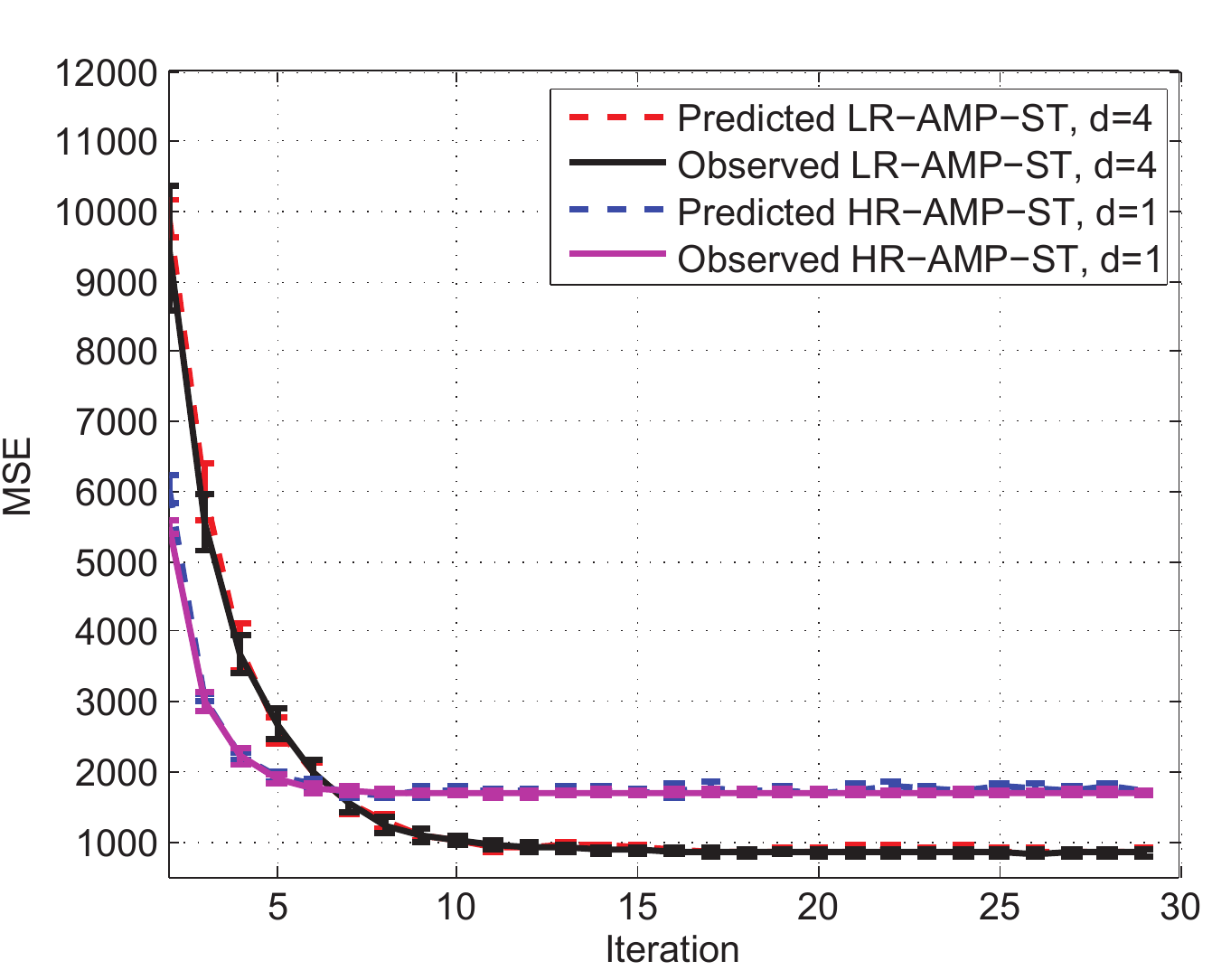}
\end{tabular}
\end{center}
\vskip -5pt
\caption{\label{SESparseAMP} Empirical intermediate MSE and predicted state evolution of HR-AMP-ST and LR-AMP-ST for image Barbara with $d=4$.}
\vskip -10pt
\end{figure}

\begin{figure*}[t]
\begin{center}
    \begin{tabular}{c@{\hspace{-0.1mm}}c}
	\includegraphics[width=2.7in]{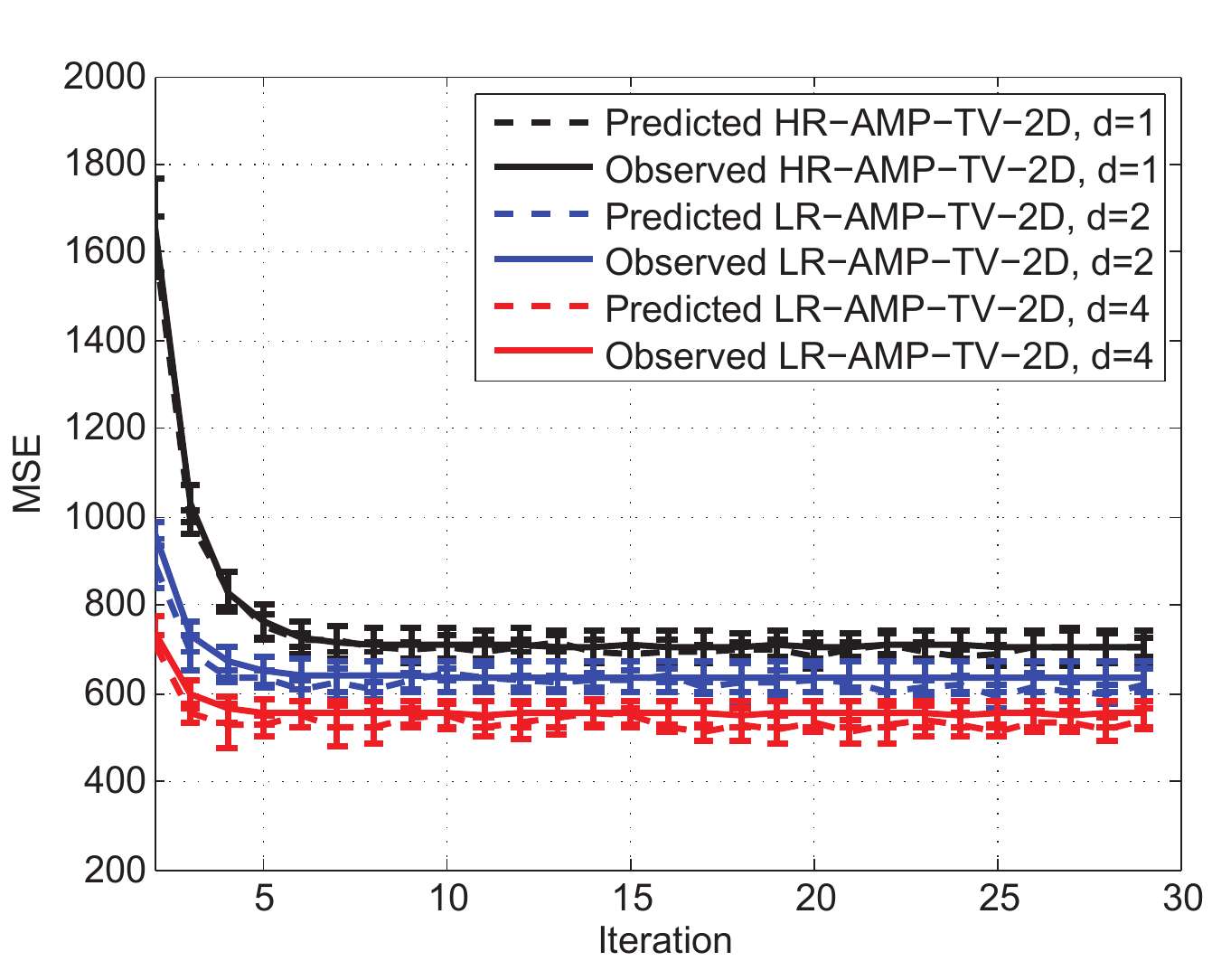} &
	\includegraphics[width=2.7in]{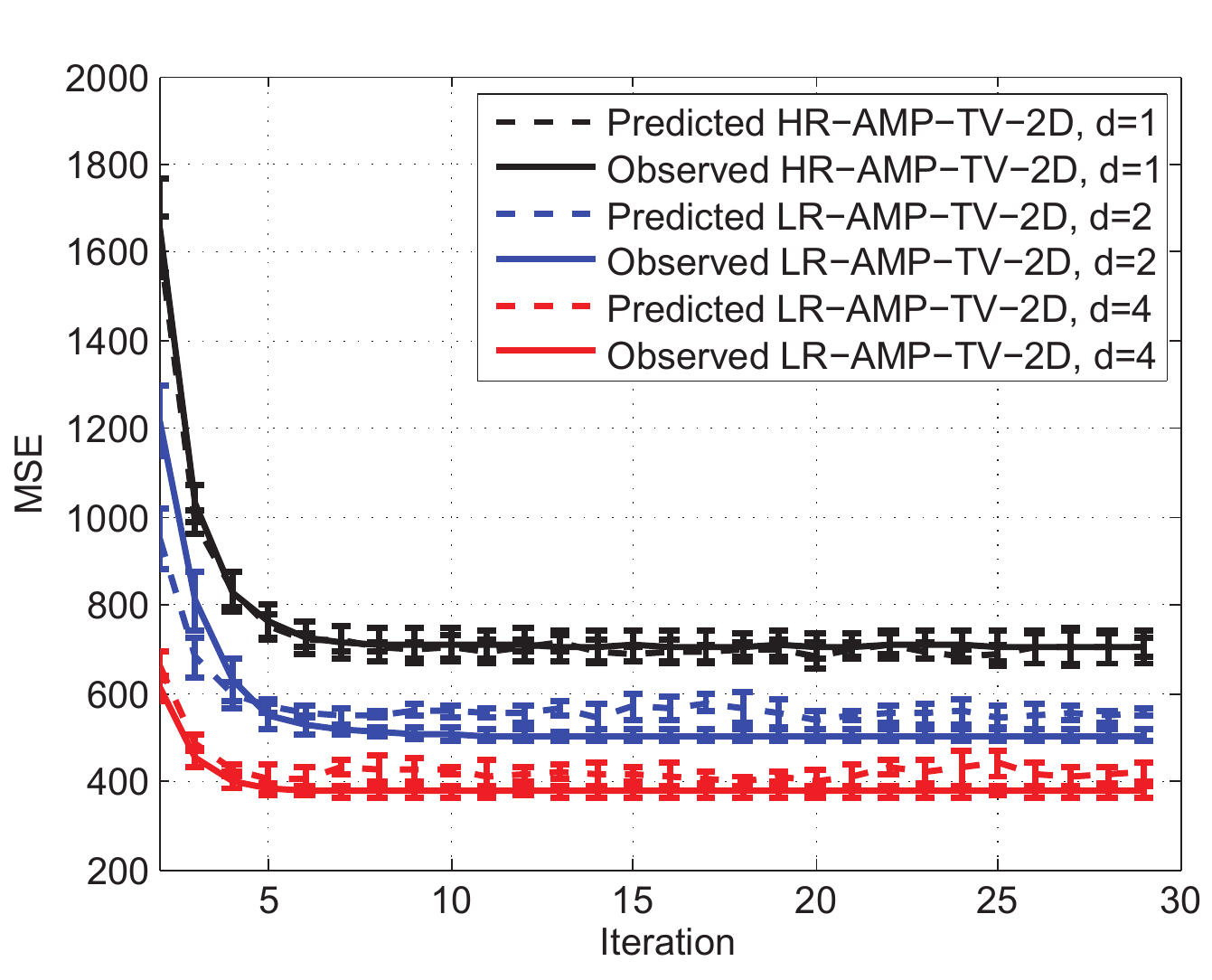} \\
(a) & (b)
\end{tabular}
\vskip -5pt
\caption{\label{SETVAMP} State evolutions of MR-AMP-TV with a CS sampling rate of ${\text{5\%}}$ and no measurement noise for the ${\text{128}} \times {\text{128}}$ Barbara image. (a) Repetition interpolator. (b) Bicubic interpolator. }
\end{center}
\vskip -15pt
\end{figure*}

\subsection{Parameter Tuning}
\label{TuningRule}

One of the main challenges in implementing different MR-AMP algorithms is the tuning of each algorithm's free parameters. Many techniques exist to estimate the noise variance in an image. In this paper, we use the following convenient feature of AMP algorithms: $\left\| {\br_d^t} \right\|_2^2/m \approx {(\sigma _d^t)^2}$ \cite{AMPthesis}.

For MR-AMP-ST, we set its threshold using three methods. For the 1D synthetic examples in Sec. \ref{sec_1DExample}, we assume the sparsity rate is known and set the thresholding parameter according to the minimax rule in \cite{AMPoriginal}. For the 2D imaging examples in Sec. \ref{sec_2DExample}, since images are not exactly sparse in transform domain, we have to estimate the sparsity rate. For sufficient large CS undersampling rate $\delta_1$ such as $10\% $ and $20\% $, we use the SURE (Stein's unbiased risk estimate)-based method in \cite{Paraless} to decide the thresholding parameter in each iteration. For very small $\delta_1$ such as $3\% $ and $4\%$, SURE does not work well since it is based on large system limit, we choose the max-min optimal threshold as determined by \cite{DAMP45}.


For AMP-TV, we use different tuning methods for 1D and 2D signals. For 1D signals, we use the source code from \cite{ssAMP} directly. For 2D images, there are many methods on how to adaptively choose the regularization parameter in TV-based image denoising, {\it e.g.}, \cite{AdapTV} and \cite{ROF}. In this paper, we use Algorithm 6 in \cite{ROF}, due to its simplicity and efficiency. In each iteration of AMP-TV-2D, a Lagrangian optimization problem is solved, whose constraint is the TV of the solution, and the Lagrangian parameter can be adaptively determined by a formula.

In AMP-ST, the Onsager term is obtained by Eq. (4.1) in \cite{AMP}. For AMP-TV-1D, the Onsager term is calculated by Eq. (5.11) in \cite{PTCPrediction}. For AMP-TV-2D, it is difficult to obtain an exact expression of the divergence. We thus apply the Monte Carlo method in \cite{DAMP} to find a good approximation of the divergence.

\subsection{State Evolution in MR-AMP}

In this part, we compare the predicted and observed performances of MR-AMP with different denoisers. Recall that the state evolution of MR-AMP is given in Eq. (\ref{SE_LR}). To compute this value, at every iteration we add white Gaussian noise with standard deviation $\sigma _d^t$ to ${\bx_{d,o}}$, denoise the signal with denoiser ${\eta _{\sigma _d^t}}(:,\tau )$, and then compute the MSE.

Fig. \ref{SESparseAMP} compares the empirical MSE and predicted state evolution of MR-AMP-ST for the test image Barbara of size $128 \times 128$, with DCT being the sparsifying basis. It can be seen that the state evolution is quite accurate. Moreover, the converged MSE per entry of the LR image is about $50\%$ smaller than that of the HR image, which verifies the motivation of this paper, {\it i.e.}, we can recover a LR signal with smaller MSE when the MSE of the HR signal is too large. Note that the LR reference image is obtained via the DCT-domain downsampling in Sec. \ref{sec_trunc}, and the corresponding MSE is the MSE between the reconstructed LR image by LR-AMP-ST and the LR reference image.

Fig. \ref{SETVAMP} shows the state evolution performance of MR-AMP-TV. Two different upsampling matrices are compared: the repetition interpolator in Eq. (\ref{EquRepetition}) (MR-AMP-TV-2D-R) and bicubic interpolator (MR-AMP-TV-2D-B). The reference LR image is obtained by Matlab's ${\text{imresize(\bx,1/d)}}$ command with bicubic interpolator. There is near perfect correspondence between the predicted and true MSEs for the repetition interpolation. For bicubic interpolator, a slight mismatch exists, because the entries of the new measurement matrix are not exactly independent. The figures also show that lower resolution provides smaller MSE, and bicubic interpolator outperforms the repetition operator.

{ Note that the denoiser in the AMP-TV-2D is essentially a non-scalar denoiser, similar to \cite{DAMP,NCSUAMP,TvampParis}. Although the state evolution for AMP with non-scalar denoisers has not been proved rigorously, the results in Fig. \ref{SETVAMP} suggest that the state evolution derived in our paper is quite accurate.}

\subsection{Performance with Synthetic 1D Signals}
\label{sec_1DExample}

In this part, we demonstrate the performance of the proposed scheme for synthetic 1D signals, which can verify the theoretical noiseless phase transition curve (PTC) and noise sensitivity.

\subsubsection{Transform Domain Approach}

To get the empirical noiseless PTC of HR-AMP-ST, we fix ${n_1}=2000$, and take $30$ equally distanced values of $\delta_1=m/n_1$ in the range of $[0.05,0.95]$, and $30$ equally distanced values of $\rho_1=k_1/m$ in $[0.05,0.95]$. For each combination of $(\delta_1, \rho_1)$, a 1D Bernoulli-Gaussian signal and its CS samples are generated before applying the HR-AMP-ST. The empirical PTCs are obtained by connecting operating points with $50\%$ success rate of the signal recovery, where the recovery is considered successful when the normalized MSE (NMSE) satisfies $\left\| {{\bx_o} - \hat \bx} \right\|_2^2/\left\| {{\bx_o}} \right\|_2^2 \leqslant {10^{ - 6}}$.

To study the empirical noiseless PTC of LR-AMP-ST, we generate a special ${n_1} \times 1$ sparse signal, whose first $n_d=n_1/d$ entries are Bernoulli-Gaussian distributed, and all other entries are $0$. According to Eq. (\ref{Model}), the truncation operator does not introduce any approximation error ${\bA}({\bI} - {\bU_d} {\bD_d})\bx$. We then run the HR-AMP-ST and LR-AMP-ST algorithms to recover the target HR and LR signals respectively. Note that we are interested in the case $m/n_1 < 1/d$, otherwise the setup is no longer a CS problem. Although the procedure of generating the HR signal here is different from that in the simulation of empirical HR-AMP-ST above, both signals belong to the same class of probability distribution if the numbers of nonzero coefficients are the same, and experimental results show that these two empirical PTCs for HR-AMP-ST coincide with each other.

\begin{figure}[tb]
\begin{center}
\begin{tabular}{c}
  \includegraphics[width=2.7in]{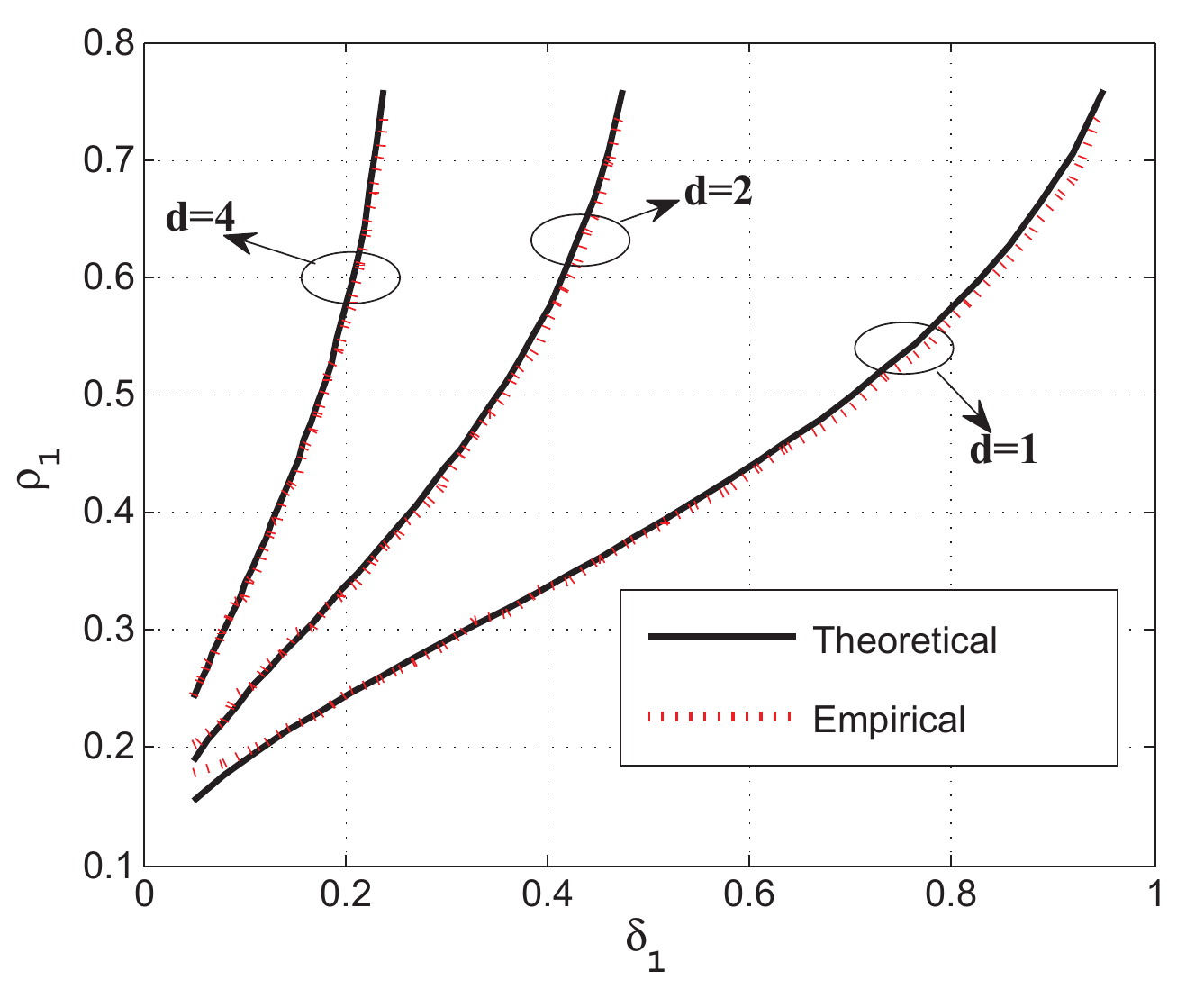}
\end{tabular}
\end{center}
\vskip -5pt
\caption{\label{PTCSparse} The theoretical and empirical PTCs of MR-AMP-ST.}
\end{figure}

The theoretical noiseless PTC in Eq. (\ref{PTCLREqu}) and the empirical noiseless PTC of LR-AMP-ST are shown in Fig. \ref{PTCSparse} for simple sparse signals with different $d$. The two sets of curves agree perfectly. It can be shown that as $d$ increases, the PTC curve shifts to the left, which means that the LR-AMP can recover the signal even when the HR-AMP fails.

\begin{table}[t]
\begin{center}
\begin{tabular}{c|c|c|c|c}
\hline
$\gamma$ & HR-AMP-ST & HR-AMP-ST & LR-AMP-ST & LR-AMP-ST \\
 & Bound & Empirical& Bound & Empirical \\
\hline
0.95 & 3.80 & 3.06 & 5.23 & 2.78 \\
0.98 & 9.80 & 7.47 & 5.23 & 4.12 \\
0.99 & 19.80 & 14.62 & 5.23 & 4.15 \\
0.998 & 39.80 & 28.89 & 5.23 & 4.79 \\
\hline
\end{tabular}
\caption{\label{NumNS}Noise sensitivity of MR-AMP-ST with $\delta_1=0.2$ and $\rho_1=0.3$. }
\end{center}
\vskip -10pt
\end{table}

The example above does not have approximation error. {Next, we construct a special case to show that the noise sensitivity of HR-AMP-ST is unbounded above the PTC, while the noise sensitivity of the LR-AMP-ST is still bounded. The setup is similar to that in \cite{AMP}, where a special 3-point distribution of $\bx$ is constructed in Lemma 4.4, whose MSE above the phase transition boundary is given by $\delta_1 \gamma / (1-\gamma)$. Therefore the MSE can go to infinity when $\gamma$ is close to 1.} We present in Table \ref{NumNS} the noisy sensitivity of MR-AMP-ST with ${n_1}=2000$, $\delta_1=0.2$, $\rho_1=0.3$ and $\sigma _w^2 = 1$. As shown in Fig. \ref{PTCSparse}, this setup is above the PTC of $d=1$, but below the PTC of $d=2$. The non-zero locations of $\bx$ are chosen with probability $1.8 \varepsilon_1$ from the first ${n_2}$ entries to generate the 3-point distribution, and with probability $0.2 \varepsilon_1$ to generate Bernoulli-Gaussian signals for the second ${n_2}$ entries, in order to fix the approximation error in Eq. (\ref{EquAMPLR}) for different $\gamma$'s. We then apply HR-AMP-ST and LR-AMP-ST to reconstruct $\bx$ and $\bx_d$.

It can be seen from Table \ref{NumNS} that as $\gamma$ approaches to $1$, the noise sensitivity bound of HR-AMP-ST keeps increasing, but the noise sensitivity bound of LR-AMP-ST is stable because all parts in Eq. (\ref{EquNSLR}) are fixed. This verifies the advantage of our LR-AMP. The empirical results of both methods are also below their noise sensitivity bounds.

\subsubsection{Spatial Domain Approach}

\begin{figure*}[tb]
\begin{center}
\begin{tabular}{c@{\hspace{-0.1mm}}c}
  \includegraphics[width=2.7in]{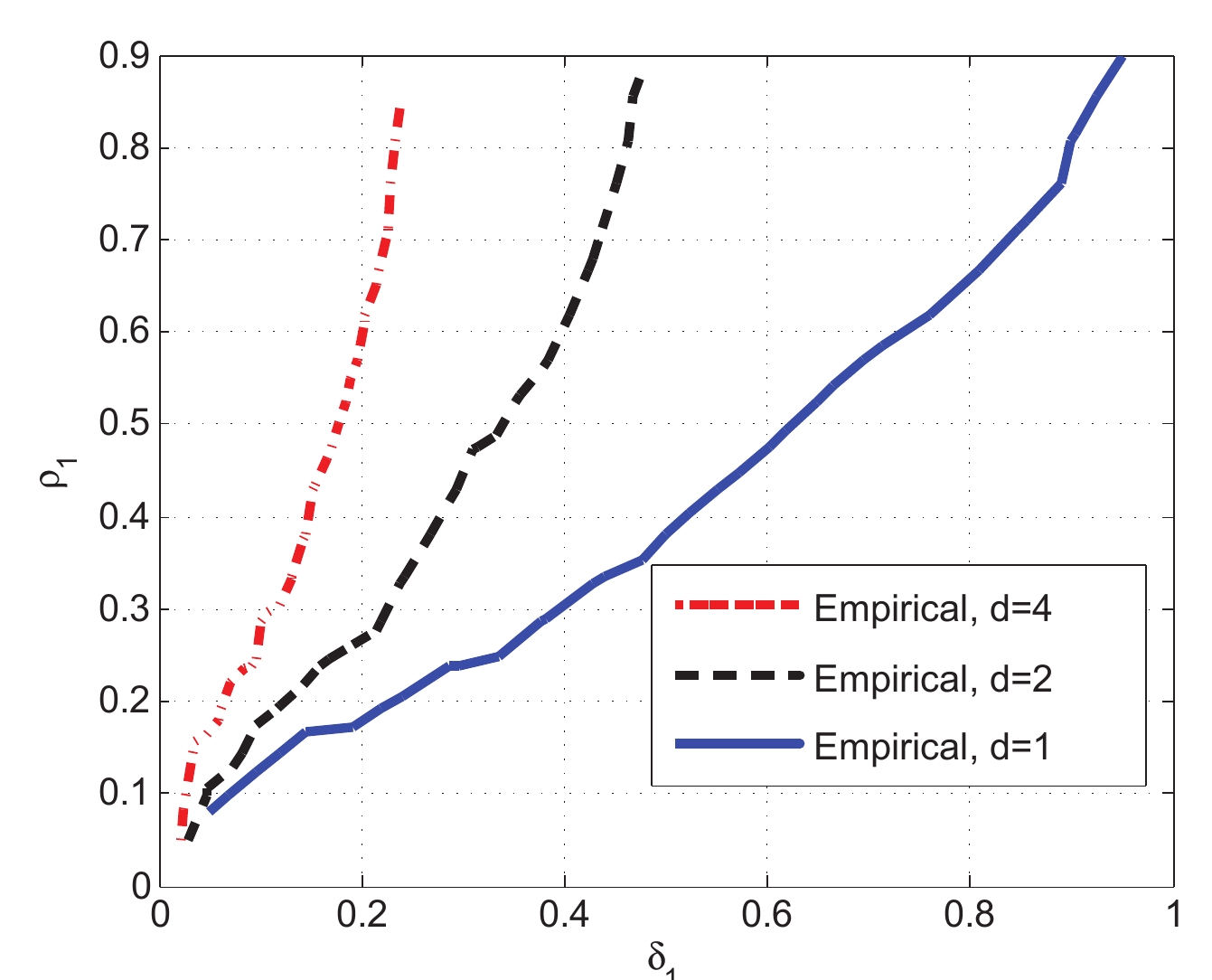} &
  \includegraphics[width=2.7in]{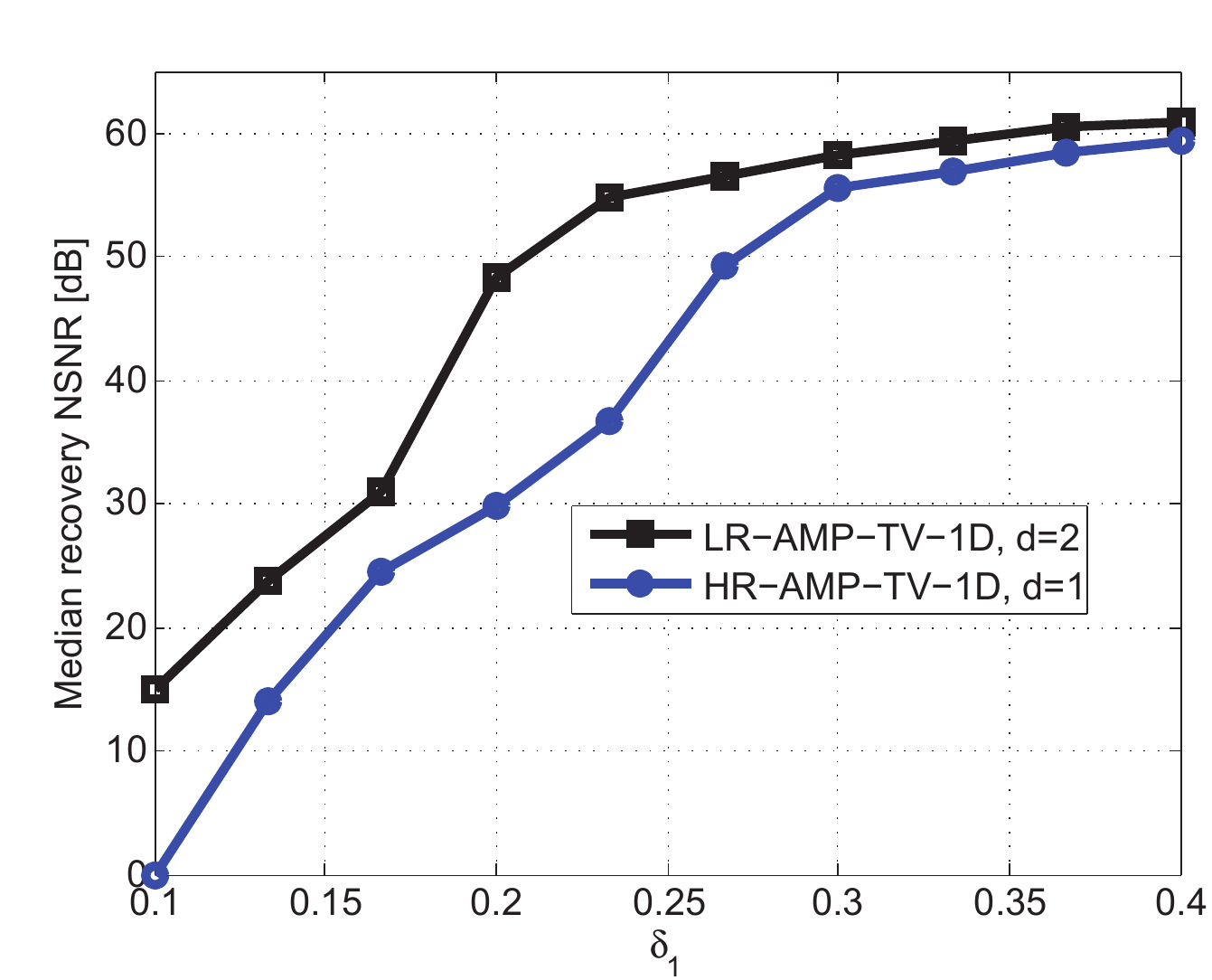} \\
  (a) & (b)
\end{tabular}
\end{center}
\vskip -5pt
\caption{\label{fig_PTCTV} (a) The empirical PTCs of MR-AMP-TV-1D for Bernoulli-Gaussian finite-difference signals. (b) MR recovery of Bernoulli-Gaussian finite-difference signals with sparsity rate ${\varepsilon_1} = 0.05$ and \text{SNR} of 60dB in the measurement. }
\vskip -10pt
\end{figure*}

It is difficult to reproduce the theoretical noiseless PTC of HR-AMP-TV-1D in \cite{PTCPrediction} since it relies on complicated numerical optimization and no open source code is available. Instead, we study the empirical noiseless PTC of HR-AMP-TV-1D by replicating an experiment from \cite{ssAMP} using its source code. We fix ${n_1}=628$, and consider a $30 \times 30$ uniform grid in the range of $\delta_1 = m/n_1 \in [0.05,0.95]$ and $\rho_1 = k_1/m \in [0.05,0.95]$. The corresponding HR Bernoulli-Gaussian 1D finite-difference signal is then generated. The empirical noiseless PTC of HR-AMP-TV-1D is shown in Fig. \ref{fig_PTCTV} (a) (with $d=1$).

To get the empirical noiseless PTCs of LR-AMP-TV-1D, we first generate LR signal ${\bx_d}$ that yields 1D Bernoulli-Gaussian finite-difference sequence with sparsity rate $d{\varepsilon_1} $. We then duplicate each entry $d$ times to get the HR piecewise constant signal with sparsity rate ${\varepsilon_1}$, according to ${\bD_d}$ and ${\bU_d}$ in Eq. (\ref{TVxd}) and (\ref{EquRepetition}). From the analysis in Sec. \ref{sec_sum}, the approximation error is zero. Successful recovery is declared when NMSE is below ${10^{ - 4}}$. The results with $d=2$ and $d=4$ are also shown in  Fig. \ref{fig_PTCTV} (a).

To study the noise sensitivity of MR-AMP-TV-1D, we recover the target HR and LR piecewise constant signals after introducing additional white Gaussian noise (AWGN) with ${\text{SNR}} \triangleq \left\| {{\bA}\bx} \right\|_2^2/{\left\| \bw \right\|_2^2} = 60dB$ in the measurement. Fig. \ref{fig_PTCTV} (b) shows the median NSNR defined as ${\text{NSNR}} \triangleq \left\| {{\bx_o}} \right\|_2^2/\left\| {{\bx_o} - \hat \bx} \right\|_2^2$ versus sampling ratio $\delta_1 = m/{n_1}$ at the fixed sparsity rate ${\varepsilon_1} = 0.05$, as in \cite{GrGAMP}. It shows that the LR-AMP-TV-1D  has lower NMSE than the HR-AMP-TV-1D. This verifies Corollary \ref{NSLR}, {\it i.e.}, the LR reconstruction has better performance than the HR one.

\begin{table*}[t]
\centering{
\vskip 10pt
\begin{tabular}{c|c|c|c|c|c|c|c|c|c}
\hline
$d$ & ${\delta _1}$ & Algorithm & Lena & Barbara & Boat  & House & Peppers & HuskerStadium & SeaWorld\\
\hline
\multirow{12}{*}{2} & \multirow{3}{*}{$5\%$} & HR-AMP-ST & 16.75 & 15.96 & 17.60 & 18.21 & 15.53 & 15.86 & 14.39 \\
& & H2L-AMP-ST & 17.40 & 16.48 & 18.30  & 18.58 & 15.93 & 16.49 & 15.10\\
& & LR-AMP-ST  & \textbf{18.02} & \textbf{17.11} & \textbf{18.77}  & \textbf{19.13} & \textbf{16.68} & \textbf{16.89} & \textbf{15.33}\\

\cline{2-10}
& \multirow{3}{*}{$10\%$} & HR-AMP-ST & 18.50 & 17.79 & 18.94  & 19.71 & 17.35 & 16.95 & 15.17\\
& & H2L-AMP-ST  & 19.43 & 18.56 & 19.97  & 20.34 & 18.19 & 18.05 & 16.10\\
& & LR-AMP-ST & \textbf{20.82} & \textbf{19.94} & \textbf{21.07} & \textbf{21.72} & \textbf{19.43} & \textbf{18.71} & \textbf{16.79}\\

\cline{2-10}
& \multirow{3}{*}{$20\%$} & HR-AMP-ST & 21.28 & 20.36 & 21.08  & 22.31 & 19.93 & 18.69 & 16.60\\
& & H2L-AMP-ST  & 22.58 & 21.69 & 22.61  & 23.46 & 21.27 & 20.13 & 18.06\\
& & LR-AMP-ST  & \textbf{24.90} & \textbf{24.25} & \textbf{24.46}  & \textbf{26.34} & \textbf{23.76} & \textbf{21.89} & \textbf{19.72}\\

\hline
\multirow{12}{*}{4} & \multirow{3}{*}{$3\%$} & HR-AMP-ST & 15.37 & 14.76 & 16.54 & 17.09 & 14.33 & 14.96 & 13.51\\
& & H2L-AMP-ST & 16.98 & 16.33 & 18.55  & 18.86 & 15.91 & \textbf{17.24} & \textbf{15.97}\\
& & LR-AMP-ST  & \textbf{18.22} & \textbf{17.66} & \textbf{18.92}  & \textbf{19.58} & \textbf{17.02} & 17.13 & 15.59\\

\cline{2-10}
& \multirow{3}{*}{$4\%$} & HR-AMP-ST & 16.03 & 15.24 & 16.95   & 17.56 & 14.87 & 15.27 & 13.75\\
& & H2L-AMP-ST  & 17.81 & 16.91 & 19.09  & 19.46 & 16.65 & \textbf{17.71} & \textbf{16.39}\\
& & LR-AMP-ST & \textbf{19.21}& \textbf{18.42} & \textbf{19.63}  & \textbf{20.31} & \textbf{17.78} & 17.66 & 15.76\\

\cline{2-10}
& \multirow{3}{*}{$5\%$} & HR-AMP-ST & 16.52 & 15.74 & 17.24  & 17.97 & 15.91 & 15.52 & 13.95\\
&& H2L-AMP-ST  & 18.44 & 17.54 & 19.57  & 20.04 & 17.29 & \textbf{18.13} & \textbf{16.72}\\
& & LR-AMP-ST  & \textbf{19.66} & \textbf{18.90} & \textbf{19.67} & \textbf{20.60} & \textbf{18.45} & 17.48 & 15.72\\

\hline
\end{tabular}
}
\vskip 5pt
\caption{\label{DCTNoiseless} PSNRs (dB) of $128 \times 128$ image reconstructions with DCT-domain MR-AMP-ST. }
\vskip -10pt
\end{table*}

\begin{table*}[t]
\centering{
\vskip 10pt
\begin{tabular}{c|c|c|c|c|c|c|c|c|c}
\hline
$d$ & ${\delta _1}$ & Algorithm & Lena & Barbara & Boat  & House & Peppers & HuskerStadium & SeaWorld\\
\hline
\multirow{12}{*}{2} & \multirow{3}{*}{$5\%$} & HR-AMP-ST & 16.58 & 15.84 & 17.81  & 17.66 & 15.31 & 15.94 & 14.33\\
& & H2L-AMP-ST & 16.85 & 16.46 & 18.55  & 18.30 & 15.83 & 16.86 & 15.04\\
& & LR-AMP-ST  & \textbf{17.35} & \textbf{17.01} & \textbf{19.13}  & \textbf{18.95} & \textbf{16.67} & \textbf{17.28} & \textbf{15.38}\\

\cline{2-10}
& \multirow{3}{*}{$10\%$} & HR-AMP-ST & 18.20 & 17.47 & 19.29  & 19.56 & 17.14 & 16.99 & 15.05\\
& & H2L-AMP-ST  & 19.13 & 18.25 & 20.53 & 20.43 & 18.05 & 18.18 & 15.98\\
& & LR-AMP-ST & \textbf{20.62} & \textbf{19.72} & \textbf{21.46}  & \textbf{21.97} & \textbf{19.46} & \textbf{19.02} & \textbf{16.79}\\

\cline{2-10}
& \multirow{3}{*}{$20\%$} & HR-AMP-ST & 21.27 & 20.15 & 21.62  & 22.68 & 20.04 & 18.86 & 16.60\\
& & H2L-AMP-ST  & 22.98 & 21.59 & 23.53  & 24.47 & 21.61 & 20.64 & 18.11\\
& & LR-AMP-ST  & \textbf{24.98} & \textbf{23.89} & \textbf{25.02}  & \textbf{26.44} & \textbf{23.51} & \textbf{22.05} & \textbf{19.80}\\

\hline
\multirow{12}{*}{4} & \multirow{3}{*}{$3\%$} & HR-AMP-ST & 15.14 & 14.67 & 16.66  & 16.41 & 14.26 & 15.01 & 13.54\\
& & H2L-AMP-ST & 16.83 & 16.40 & 18.90  & 18.31 & 15.95 & \textbf{17.45} & \textbf{16.19}\\
& & LR-AMP-ST  & \textbf{17.83} & \textbf{17.24} & \textbf{19.31}  & \textbf{19.28} & \textbf{16.93} & 17.33 & 15.52\\

\cline{2-10}
& \multirow{3}{*}{$4\%$} & HR-AMP-ST & 15.62 & 15.16 & 17.09  & 17.08 & 14.70 & 15.40 & 13.69\\
& & H2L-AMP-ST  & 17.47 & 17.04 & 19.53  & 19.21 & 16.63 & \textbf{18.07} & \textbf{16.50} \\
& & LR-AMP-ST & \textbf{18.73} & \textbf{18.09} & \textbf{19.70}  & \textbf{19.78} & \textbf{17.60} & 17.74 & 15.78\\

\cline{2-10}
& \multirow{3}{*}{$5\%$} & HR-AMP-ST & 15.99 & 15.58 & 17.50  & 17.52 & 15.19 & 15.66 & 13.90\\
&& H2L-AMP-ST  & 17.99 & 17.63 & \textbf{20.18}  & 19.81 & 17.25 & \textbf{18.51} & \textbf{16.87}\\
& & LR-AMP-ST  & \textbf{19.00} & \textbf{18.47} & 19.65  & \textbf{20.14} & \textbf{17.84} & 17.55 & 15.60\\

\hline
\end{tabular}
}
\vskip 5pt
\caption{\label{WaveletNoiseless} PSNRs (dB) of $128 \times 128$ image reconstructions with wavelet-domain MR-AMP-ST. }
\vskip -10pt
\end{table*}

\begin{table*}[t]
\centering{
\vskip 10pt
\begin{tabular}{c|c|c|c|c|c|c|c|c|c}
\hline
$d$ & ${\delta _1}$ & Algorithm & Lena & Barbara & Boat & House & Peppers & HuskerStadium & SeaWorld \\
\hline
\multirow{12}{*}{2} & \multirow{4}{*}{$5\%$} & HR-AMP-TV-2D & 20.88 & 19.66 & 20.67  & 22.85 & 19.60 & 18.36 & 16.13\\
& & H2L-AMP-TV-2D & 22.55 & 21.09 & 22.51  & 24.53 & 21.07 & 20.27 & \textbf{17.93}\\
& & LR-AMP-TV-2D-R  & 21.79  & 20.25 & 22.17  & 23.80 & 20.13 & 19.95 & 17.75\\
& & LR-AMP-TV-2D-B & \textbf{22.68} & \textbf{21.17} & \textbf{22.67}  & \textbf{25.07} & \textbf{21.13} & \textbf{20.33} & 17.87\\

\cline{2-10}
& \multirow{4}{*}{$10\%$} & HR-AMP-TV-2D & 23.62 & 22.18 & 22.80  & 26.55 & 22.51 & 20.20 & 17.64\\
& & H2L-AMP-TV-2D  & \textbf{25.83} & 24.08 & \textbf{25.35}  & \textbf{28.91} & \textbf{24.63} & \textbf{22.70} & \textbf{19.97}\\
& & LR-AMP-TV-2D-R & 23.83 & 22.32 & 24.07  & 26.42 & 22.31 & 21.64 & 19.16\\
& & LR-AMP-TV-2D-B & 25.66 & \textbf{24.18} & 25.24  & 28.82 & 24.33 & 22.63 & 19.95\\

\cline{2-10}
& \multirow{4}{*}{$20\%$} & HR-AMP-TV-2D & 26.51 & 25.05 & 25.02  & 30.91 & 25.70 & 22.15 & 19.31\\
& & H2L-AMP-TV-2D  & \textbf{29.49} & \textbf{27.65} & \textbf{28.39}  & \textbf{34.11} & \textbf{28.49} & \textbf{25.38} & 22.26\\
& & LR-AMP-TV-2D-R  & 26.24 & 24.83 & 26.19  & 29.23 & 24.60 & 23.60 & 20.85\\
& & LR-AMP-TV-2D-B & 28.92 & 27.63 & 27.99  & 32.44 & 27.51 & 25.28 & \textbf{22.34}\\

\hline
\multirow{12}{*}{4} & \multirow{4}{*}{$3\%$} & HR-AMP-TV-2D & 18.69 & 17.90 & 18.89  & 20.32 & 17.71 & 16.77 & 15.08 \\
& & H2L-AMP-TV-2D & 21.75 & \textbf{20.80} & 22.29  & 23.55 & \textbf{20.70} & 20.26 & 18.71\\
& & LR-AMP-TV-2D-R  & 20.73 & 19.63 & 22.10  & 23.59 & 19.33 & 20.22 & 18.80\\
& & LR-AMP-TV-2D-B & \textbf{21.95} & 20.49 & \textbf{23.02}  & \textbf{24.47} & 20.46 & \textbf{20.92} & \textbf{19.22}\\

\cline{2-10}
& \multirow{4}{*}{$4\%$} & HR-AMP-TV-2D & 19.89 & 18.89 & 19.84  & 21.64 & 18.83 & 17.66 & 15.65\\
& & H2L-AMP-TV-2D  & \textbf{23.43} & \textbf{22.21} & 23.77  & 25.39 & \textbf{22.26} & 21.62 & 19.75\\
& & LR-AMP-TV-2D-R & 21.53 & 20.28 & 22.75  & 24.19 & 20.31 & 20.81 & 19.35\\
& & LR-AMP-TV-2D-B & 22.99 & 21.54 & \textbf{23.93}  & \textbf{25.55} & 21.66 & \textbf{21.64} & \textbf{19.93}\\

\cline{2-10}
& \multirow{4}{*}{$5\%$} & HR-AMP-TV-2D & 20.88 & 19.66 & 20.67  & 22.85 & 19.60 & 18.36 & 16.13\\
& & H2L-AMP-TV-2D  & \textbf{24.86} & \textbf{23.24} & \textbf{25.07}  & \textbf{27.05} & \textbf{23.40} & \textbf{22.74} & \textbf{20.49}\\
& & LR-AMP-TV-2D-R  & 22.24 & 20.86 & 23.41  & 25.09 & 20.75 & 21.51 & 19.66\\
& & LR-AMP-TV-2D-B & 23.97 & 22.32 & 24.80  & 26.58 & 22.52 & 22.54 & 20.30\\

\hline
\end{tabular}
}
\vskip 5pt
\caption{\label{TVNoiseless} PSNRs (dB) of $128 \times 128$ image reconstructions with spatial-domain MR-AMP-TV-2D. }
\vskip -10pt
\end{table*}

\begin{figure*}[tb]
\begin{center}
\begin{tabular}{@{\hspace{-9mm}}c@{\hspace{-8.5mm}}c@{\hspace{-13mm}}c@{\hspace{-13mm}}c@{\hspace{-8.5mm}}c@{\hspace{-13mm}}c}
    \includegraphics[width=1.7in]{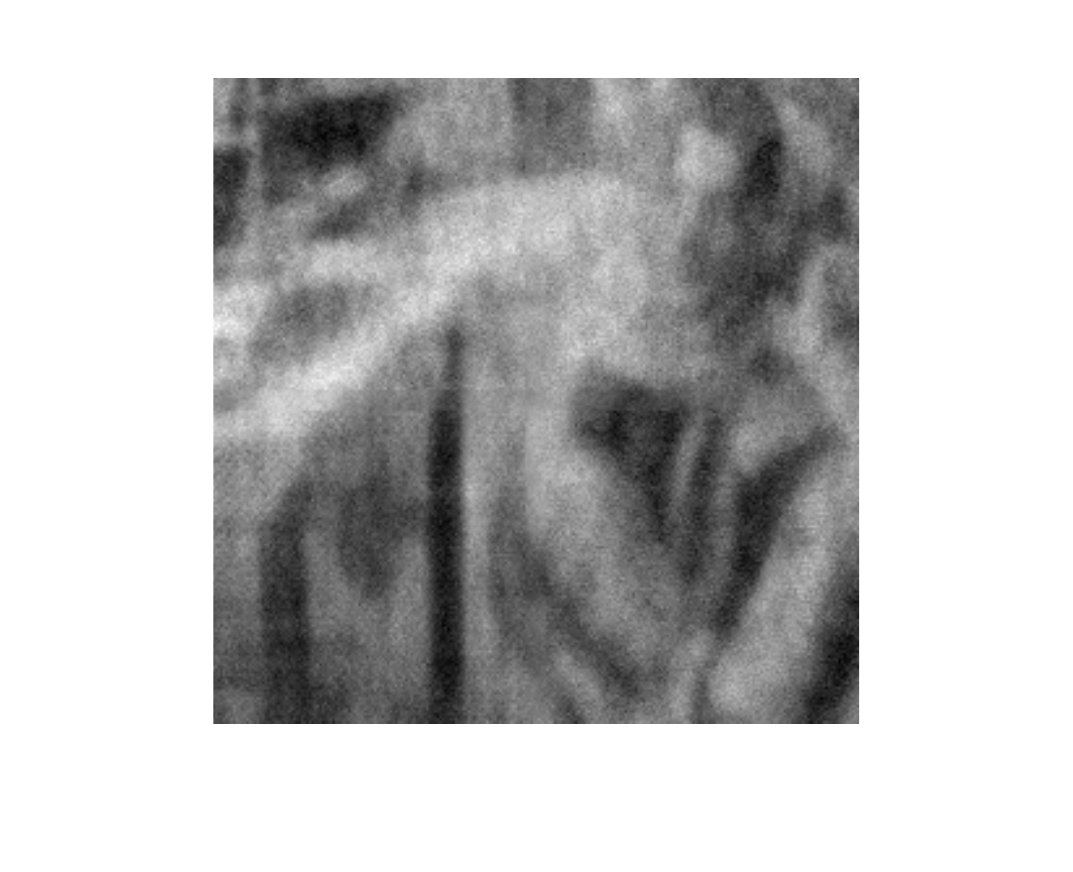} &
    \includegraphics[width=1.7in]{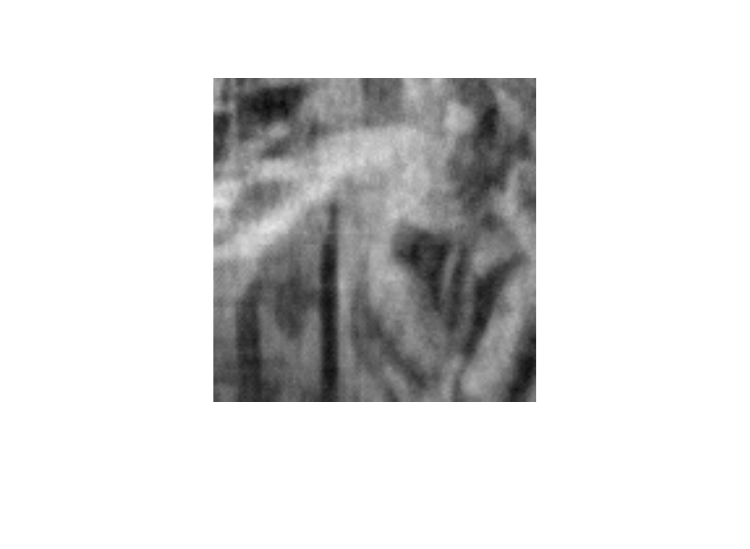} &
    \includegraphics[width=1.7in]{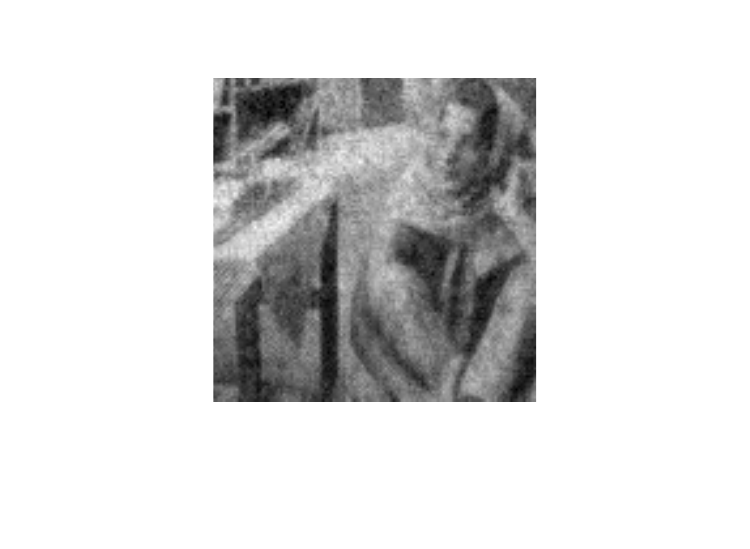} &
     \includegraphics[width=1.7in]{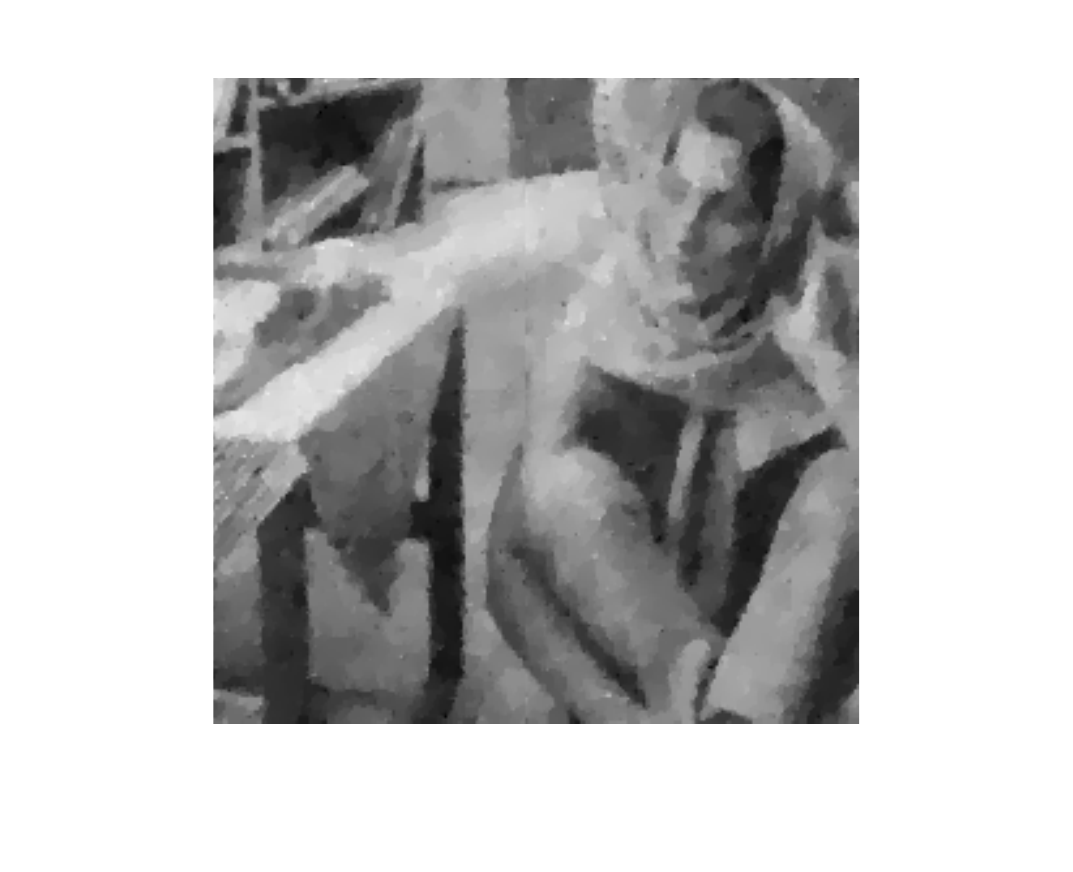} &
    \includegraphics[width=1.7in]{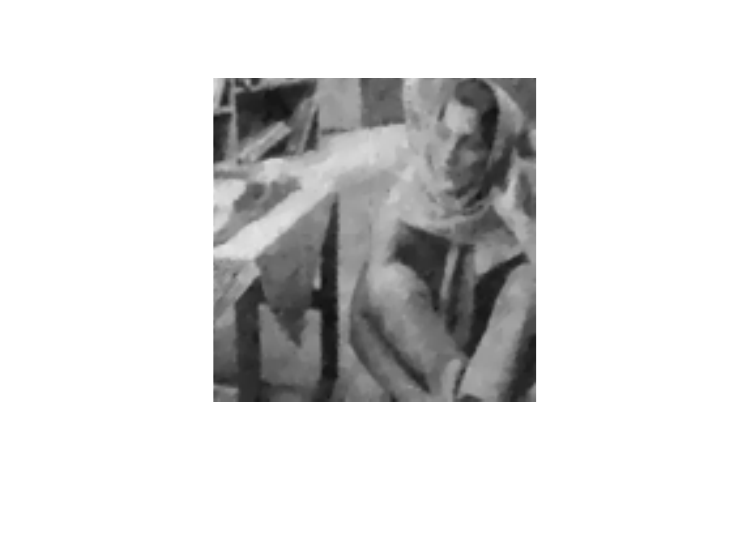} &
    \includegraphics[width=1.7in]{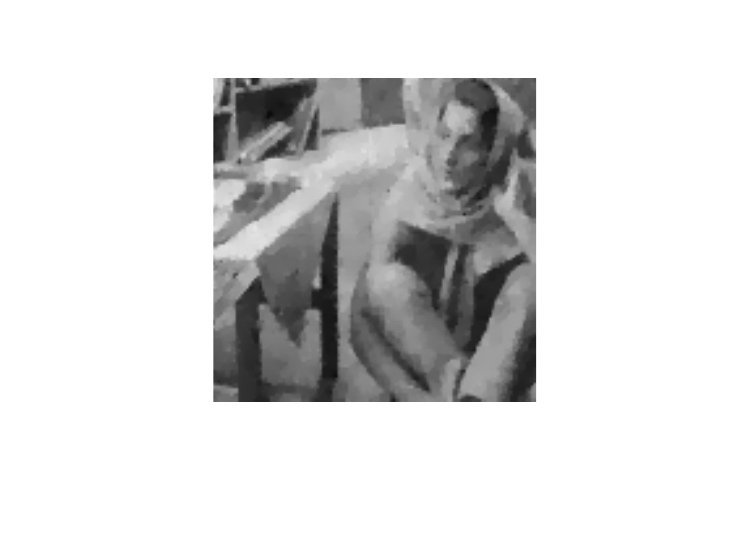} \\
(a) & (b) & (c) & (d) & (e) & (f)\\

\end{tabular}
\vskip 5pt
\end{center}
\vskip -5pt
\caption{\label{Visual} Reconstruction of $10\%$ sampled $256 \times 256$ Barbara image with  dowansampling factor $d=2$ and DCT as the sparsifying basis for MR-AMP-ST.
 (a) HR-AMP-ST (20.32dB). (b) H2L-AMP-ST (21.31dB). (c) LR-AMP-ST (22.72dB). (d) HR-AMP-TV-2D (25.06dB). (e) H2L-AMP-TV-2D  (27.75dB). (f) LR-AMP-TV-2D-B (27.54dB).}
\end{figure*}

\begin{figure*}[tb]
\begin{center}
\begin{tabular}{@{\hspace{-9mm}}c@{\hspace{-8.5mm}}c@{\hspace{-13mm}}c@{\hspace{-13mm}}c@{\hspace{-8.5mm}}c@{\hspace{-13mm}}c}
    \includegraphics[width=1.7in]{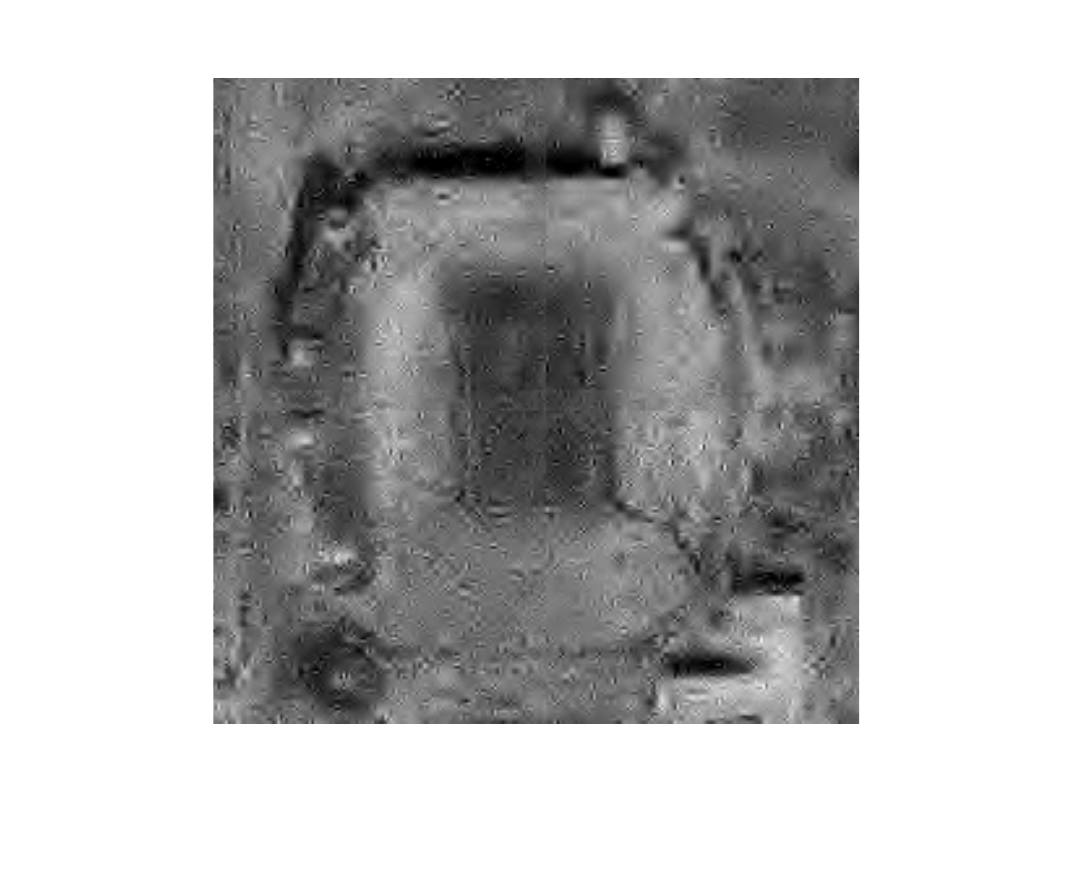} &
    \includegraphics[width=1.7in]{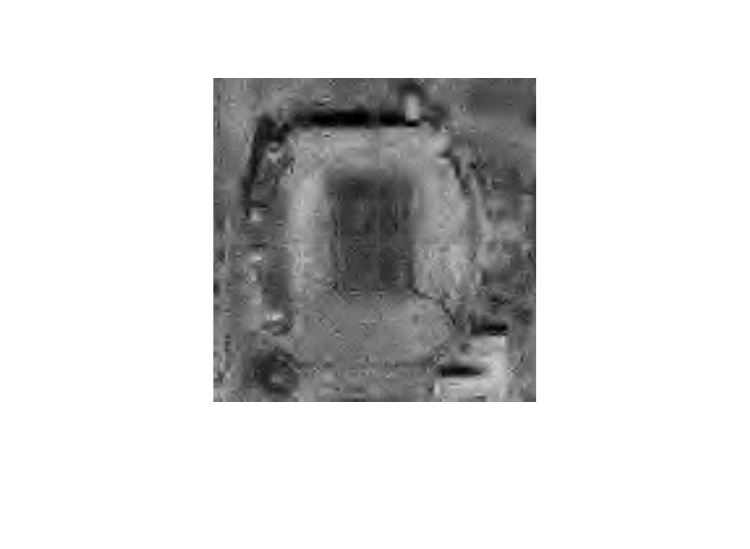} &
    \includegraphics[width=1.7in]{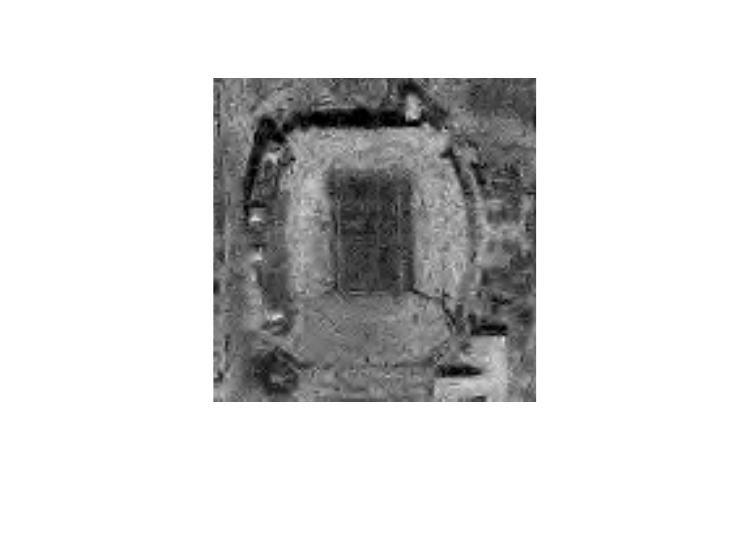} &
    \includegraphics[width=1.7in]{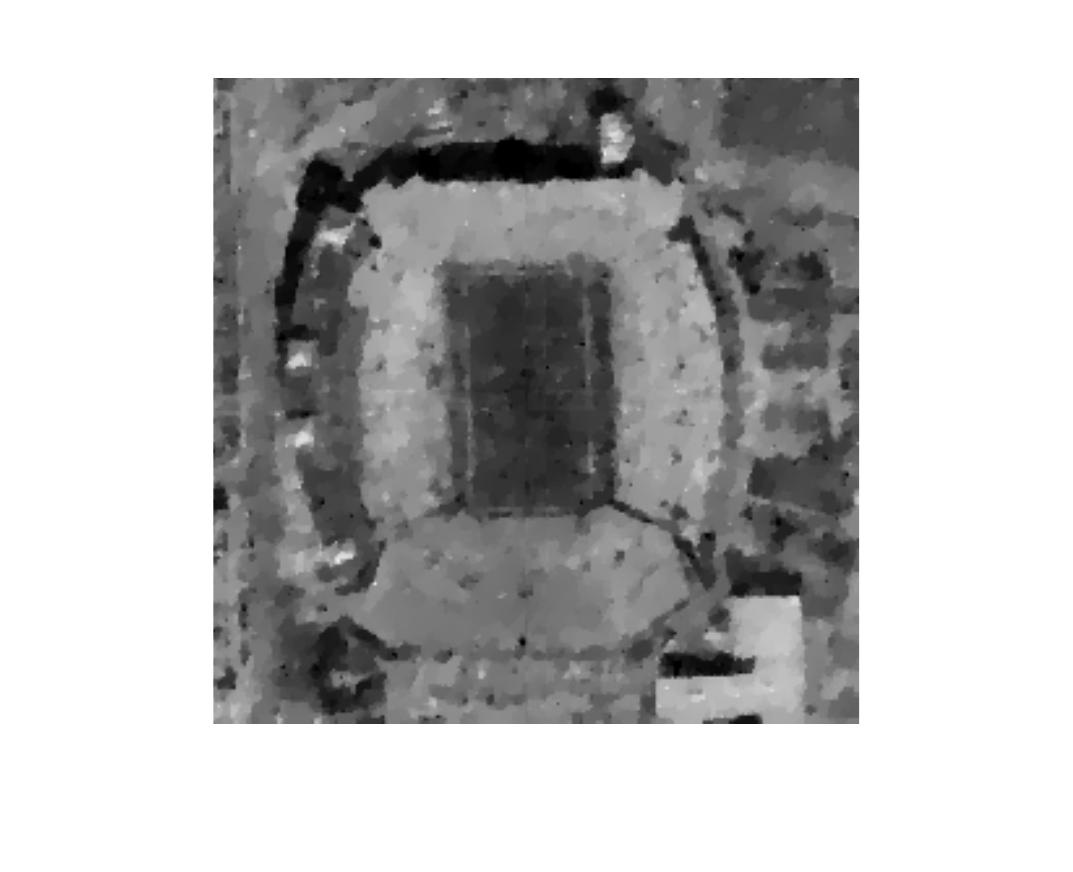} &
    \includegraphics[width=1.7in]{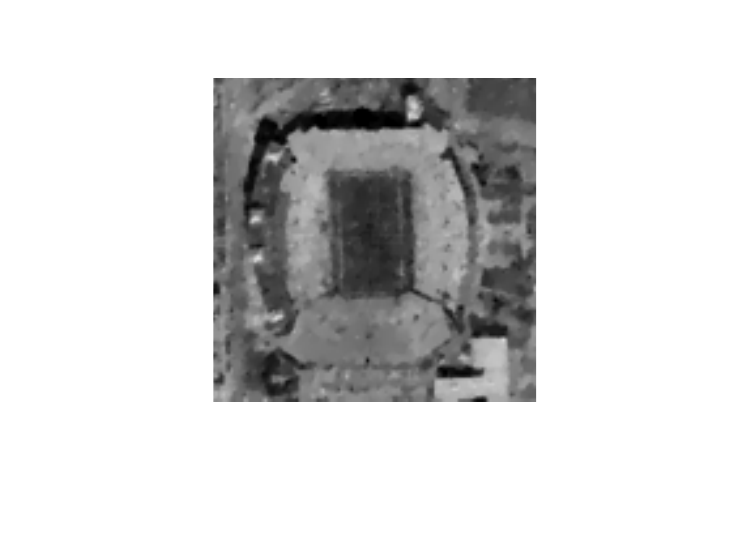} &
    \includegraphics[width=1.7in]{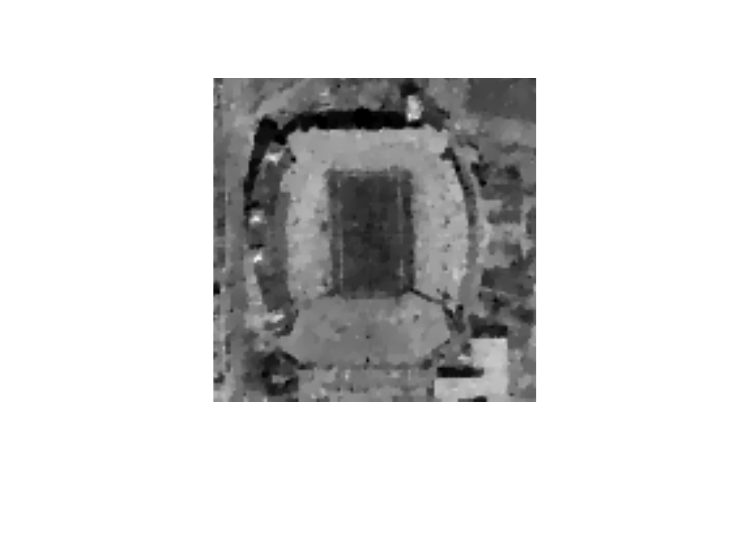} \\
(a) & (b) & (c) & (d) & (e) & (f)\\
\end{tabular}
\vskip 5pt
\end{center}
\vskip -5pt
\caption{\label{HuskerStadium} Reconstructed of $20\%$ sampled $256 \times 256$ HuskerStadium image with downsampling factor $d=2$ and D8 wavelet as the sparsifying basis for MR-AMP-ST.
 (a) HR-AMP-ST (18.65dB). (b) H2L-AMP-ST (20.22dB). (c) LR-AMP-ST (21.05dB). (d) HR-AMP-TV-2D (21.66dB). (e) H2L-AMP-TV-2D (24.52dB). (f) LR-AMP-TV-2D-B (24.38dB) .}
\end{figure*}

\begin{table*}[t]
\centering{
\vskip 10pt
\begin{tabular}{c|c|c|c|c|c|c|c|c|c}
\hline
$d$ & ${\delta _1}$ & Algorithm & Lena & Barbara & Boat & House & Peppers & HuskerStadium & SeaWorld \\
\hline
\multirow{8}{*}{2} & \multirow{4}{*}{$10\%$} & HR-AMP-TV-2D & \textbf{23.62} & \textbf{22.18} & 22.80  & \textbf{26.55} & \textbf{22.51} & 20.20 & 17.64\\
& & HR-TVAL3  & 23.36 & 21.80 & \textbf{22.94} & 26.21 & 21.89 & \textbf{20.32} & \textbf{17.71}\\
\cline{3-10}
& & LR-AMP-TV-2D-B & \textbf{25.66}  & \textbf{24.18} & \textbf{25.24}  & \textbf{28.82} & \textbf{24.33} & \textbf{22.63} & \textbf{19.95}\\
& & LR-TVAL3 & 25.44 & 24.05 & 25.09  & 28.60 & 23.98 & 21.94 & 19.24\\

\cline{2-10}
& \multirow{4}{*}{$20\%$} & HR-AMP-TV-2D & 26.51 & 25.05 & 25.02  & 30.91 & \textbf{25.70} & 22.15 & 19.31\\
& & HR-TVAL3  & \textbf{26.80} & \textbf{25.22} & \textbf{25.49} & \textbf{31.79} & 25.65 & \textbf{22.42} & \textbf{19.62}\\
\cline{3-10}
& & LR-AMP-TV-2D-B  & \textbf{28.92} & \textbf{27.63} & \textbf{27.99}  & \textbf{32.44} & \textbf{27.51} & \textbf{25.28} & \textbf{22.34}\\
& & LR-TVAL3 & 28.58 & 27.45 & 27.59  & 31.67 & 27.02 & 24.19 & 21.38\\

\hline

\multirow{8}{*}{4} & \multirow{4}{*}{$4\%$} & HR-AMP-TV-2D & \textbf{19.89} & \textbf{18.89} & 19.84  & \textbf{21.64} & \textbf{18.83} & 17.66 & 15.65\\
& & HR-TVAL3  & 19.69 & 18.77 & \textbf{20.21}  & 21.43 & 18.48 & \textbf{18.21} & \textbf{16.02}\\
\cline{3-10}
& & LR-AMP-TV-2D-B & \textbf{22.99} & \textbf{21.54} & \textbf{23.93}  & \textbf{25.55} & \textbf{21.66} & \textbf{21.64} & \textbf{19.93}\\
& & LR-TVAL3 & 22.77 & 21.31 & 23.85  & 25.19 & 21.56 & 20.95 & 19.80\\

\cline{2-10}
& \multirow{4}{*}{$5\%$} & HR-AMP-TV-2D & \textbf{20.88} & \textbf{19.66} & 20.67  & \textbf{22.85} & \textbf{19.60} & 18.36 & 16.13\\
& & HR-TVAL3  & 20.58 & 19.29 & \textbf{20.94}  & 22.57 & 19.15 & \textbf{18.70} & \textbf{16.38}\\
\cline{3-10}
& & LR-AMP-TV-2D-B  & \textbf{23.97} & \textbf{22.32} & \textbf{24.80}  & \textbf{26.58} & \textbf{22.52} & \textbf{22.54} & 20.30\\
& & LR-TVAL3 & 23.80 & 21.85 & 24.40  & 26.15 & 22.46 & 21.45 & \textbf{20.45}\\

\hline
\end{tabular}
}
\vskip 5pt
\caption{\label{Compare1} Comparison of the final reconstruction results in PSNR between TVAL3 and AMP-TV-2D.}
\vskip -10pt
\end{table*}

%
%
%

\begin{table*}[tb]
\centering{
\vskip 10pt
\scalebox{0.95}{
\begin{tabular}{c|c|c|c|c|c|c|c|c|c|c}
\hline
\multicolumn{11}{c}{AWGN with standard deviation 20} \\
\hline
\multirow{4}{*}{DCT} & \multirow{4}{*}{d = 2} & ${\delta _1}$ & ${5\%}$ & ${10\%}$ & ${15\%}$ & \multirow{4}{*}{d=4} & ${\delta _1}$ & ${3\%}$ & ${4\%}$ & ${5\%}$  \\
\cline{3-6}
\cline{8-11}
& & HR-AMP-ST & 16.00 & 17.70 & 19.92 & {} & HR-AMP-ST & 14.74 & 15.22 & 15.73  \\
& & H2L-AMP-ST & 16.45 & 18.43 & 21.08 & {}& H2L-AMP-ST & 16.32 & 16.90 & 17.53 \\
& & LR-AMP-ST & \textbf{17.12} & \textbf{19.65} & \textbf{22.88} & {} & LR-AMP-ST & \textbf{17.56} & \textbf{18.37} & \textbf{18.71} \\

\hline
\multirow{4}{*}{Wavelet} & \multirow{4}{*}{d=2} & ${\delta _1}$ & ${5\%}$ & ${10\%}$ & ${15\%}$ & \multirow{4}{*}{d=4} & ${\delta _1}$ & ${3\%}$ & ${4\%}$ & ${5\%}$  \\
\cline{3-6}
\cline{8-11}
& & HR-AMP-ST & 15.80 & 17.47 & 19.64 & & HR-AMP-ST & 14.65 & 15.14 & 15.56  \\
& & H2L-AMP-ST & 16.44 & 18.33 & 21.02 & & H2L-AMP-ST & 16.38 & 17.04 & 17.61 \\
& & LR-AMP-ST & \textbf{16.85} & \textbf{19.56} & \textbf{22.79} & & LR-AMP-ST & \textbf{17.20} & \textbf{17.98} & \textbf{18.29} \\

\hline
\multirow{5}{*}{TV} & \multirow{5}{*}{d=2} & ${\delta _1}$ & ${5\%}$ & ${10\%}$ & ${15\%}$ & \multirow{5}{*}{d=4} & ${\delta _1}$ & ${3\%}$ & ${4\%}$ & ${5\%}$  \\
\cline{3-6}
\cline{8-11}
& & HR-AMP-TV-2D & 19.59 & 21.84 & 23.93 & & HR-AMP-TV-2D & 17.82 & 18.85 & 19.59  \\
& & H2L-AMP-TV-2D & 21.00 & 23.68 & 26.24 & & H2L-AMP-TV-2D & \textbf{20.72} & \textbf{22.13} & \textbf{23.09} \\
& & LR-AMP-TV-2D-R & 20.20 & 22.08 & 24.17 & & LR-AMP-TV-2D-R & 19.60 & 20.25 & 20.80 \\
& & LR-AMP-TV-2D-B & \textbf{21.05} & \textbf{23.70} & \textbf{26.31} & & LR-AMP-TV-2D-B & 20.45 & 21.48 & 22.22 \\

\hline
\multicolumn{11}{c}{AWGN with standard deviation 40} \\
\hline
\multirow{4}{*}{DCT} & \multirow{4}{*}{d = 2} & ${\delta _1}$ & ${5\%}$ & ${10\%}$ & ${15\%}$ & \multirow{4}{*}{d=4} & ${\delta _1}$ & ${3\%}$ & ${4\%}$ & ${5\%}$  \\
\cline{3-6}
\cline{8-11}
& & HR-AMP-ST & 15.89 & 17.37 & 19.06 & {} & HR-AMP-ST & 14.67 & 15.16 & 15.59  \\
& & H2L-AMP-ST & 16.34 & 18.03 & 19.94 & {}& H2L-AMP-ST & 16.26 & 16.96 & 17.43 \\
& & LR-AMP-ST & \textbf{17.00} & \textbf{19.11} & \textbf{21.21} & {} & LR-AMP-ST & \textbf{17.41} & \textbf{18.09} & \textbf{18.24} \\

\hline
\multirow{4}{*}{Wavelet} & \multirow{4}{*}{d=2} & ${\delta _1}$ & ${5\%}$ & ${10\%}$ & ${15\%}$ & \multirow{4}{*}{d=4} & ${\delta _1}$ & ${3\%}$ & ${4\%}$ & ${5\%}$  \\
\cline{3-6}
\cline{8-11}
& & HR-AMP-ST & 15.68 & 17.19 & 18.72 & & HR-AMP-ST & 14.61 & 15.08 & 15.42  \\
& & H2L-AMP-ST & 16.21 & 17.87 & 19.75 & & H2L-AMP-ST & 16.37 & 17.00 & 17.49 \\
& & LR-AMP-ST & \textbf{16.81} & \textbf{18.95} & \textbf{21.10} & & LR-AMP-ST & \textbf{17.00} & \textbf{17.76} & \textbf{17.93} \\

\hline
\multirow{5}{*}{TV} & \multirow{5}{*}{d=2} & ${\delta _1}$ & ${5\%}$ & ${10\%}$ & ${15\%}$ & \multirow{5}{*}{d=4} & ${\delta _1}$ & ${3\%}$ & ${4\%}$ & ${5\%}$  \\
\cline{3-6}
\cline{8-11}
& & HR-AMP-TV-2D & 19.36 & 21.13 & 22.52 & & HR-AMP-TV-2D & 17.75 & 18.70 & 19.36  \\
& & H2L-AMP-TV-2D & 20.70 & 22.80 & \textbf{24.51} & & H2L-AMP-TV-2D & \textbf{20.62} & \textbf{21.94} & \textbf{22.75} \\
& & LR-AMP-TV-2D-R & 20.02 & 21.58 & 23.05 & & LR-AMP-TV-2D-R & 19.55 & 20.16 & 20.61 \\
& & LR-AMP-TV-2D-B & \textbf{20.73} & \textbf{22.81} & 24.44 & & LR-AMP-TV-2D-B & 20.35 & 21.31 & 21.91 \\
\hline
\end{tabular}
}
\vskip 15pt
\caption{\label{NoisyResults} PSNRs (dB) of reconstruction of $128 \times 128$ Barbara image with varying amounts of additive Gaussian measurement noise.}
}
\end{table*}

\begin{table*}[t]
\centering{
\vskip 10pt
\begin{tabular}{c|c|c|c|c|c|c|c|c|c}
\hline
$d$ & ${\delta _1}$ & Algorithm & Lena & Barbara & Boat & House & Peppers & HuskerStadium & SeaWorld \\
\hline
\multirow{4}{*}{2} & \multirow{4}{*}{$10\% $} & HR-AMP-ST & 18.50 & 17.79 & 18.94  & 19.71 & 17.35 & 16.95 & 15.17\\
& & L2H-AMP-ST & \textbf{20.14} & \textbf{19.46} & \textbf{20.18} & \textbf{21.25} & \textbf{19.02} & \textbf{18.03} & \textbf{16.07}\\
\cline{3-10}
&  & HR-AMP-TV-2D-B  & 23.62  & 22.18 & 22.80  & \textbf{26.55} & \textbf{22.51} & 20.20 & 17.64\\
& & L2H-AMP-TV-2D-B & \textbf{23.65} & \textbf{22.52} & \textbf{22.86}  & {26.21} & {22.30} & \textbf{20.25} & \textbf{17.79}\\
\hline
\multirow{4}{*}{4} & \multirow{4}{*}{$5\% $} & HR-AMP-ST & 16.52 & 15.74 & 17.24  & 17.97 & 15.91 & 15.52 & \textbf{13.95}\\
& & L2H-AMP-ST & \textbf{18.76} & \textbf{17.85} & \textbf{19.05} & \textbf{20.00} & \textbf{17.65} & \textbf{16.38} & 11.93\\
\cline{3-10}
&  & HR-AMP-TV-2D-B  & \textbf{20.88}  & \textbf{19.66} & \textbf{20.67}  & \textbf{22.85} & \textbf{19.60} & \textbf{18.36} & \textbf{16.13}\\
& & L2H-AMP-TV-2D-B & {20.47} & {19.39} & {20.56}  & {22.25} & {19.10} & {18.23} & {16.11}\\

\hline
\end{tabular}
}
\vskip 5pt
\caption{\label{L2H} PSNRs (dB) of $128 \times 128$ image reconstructions with HR-AMP and L2H-AMP. The transform domain in AMP-ST is DCT. }
\vskip -10pt
\end{table*}

\begin{table*}[tb]
\begin{center}
\begin{tabular}{c|c|c|c|c|c|c}
\hline
\multicolumn{3}{c|}{d=2 for LR-AMP-ST} & \multicolumn{4}{c}{d=2 for LR-AMP-TV-2D} \\
\hline
$\delta_1 \%$ & HR-AMP-ST & LR-AMP-ST & $\delta_1\%$ & HR-AMP-TV-2D & LR-AMP-TV-2D-R & LR-AMP-TV-2D-B \\
\hline
$5$ & 10.8969 & 3.7318 & $5$ & 9.9753 & 2.4964 & 2.3032\\
$10$ & 11.5907 & 3.9249 & $10$ & 6.9049 & 2.2856 & 2.0812\\
$20$ & 12.5869 & 4.1898 & $20$ & 5.7327 & 2.6401 & 2.4869\\
\hline
\multicolumn{3}{c|}{d=4 for LR-AMP-ST} & \multicolumn{4}{c}{d=4 for LR-AMP-TV-2D} \\
\hline
$\delta_1 \%$ & HR-AMP-ST & LR-AMP-ST & $\delta_1\%$ & HR-AMP-TV-2D & LR-AMP-TV-2D-R & LR-AMP-TV-2D-B \\
\hline
$3$ & 0.2489 & 0.0075 & $3$ & 14.4794 & 0.8791 & 0.8486 \\
$4$ & 0.3205 & 0.0080 & $4$ & 11.8068 & 0.8831 & 0.8594 \\
$5$ & 0.3937 & 0.0107 & $5$ & 9.9753 & 0.9104 & 0.9005 \\
\hline
\end{tabular}
\caption{\label{RunTime}
CPU running time in seconds of different methods for the $128 \times 128$ Barbara image.}
\end{center}
\vskip -10pt
\end{table*}

\subsection{Performance with 2D Images}
\label{sec_2DExample}

In this part, we apply the MR-AMP theory to MR 2D image reconstruction. All reported experimental results are the averages of 20 Monte Carlo simulations.

\subsubsection{Target LR image}

The target LR images are different when different downsampling matrices are used. For the transform-domain approach, the target LR image ${\bX_d}$ is represented by Eq. (\ref{EqUpDown2D}). Both DCT and the Daubechies-8 (D8) wavelet are tested. For the spatial-domain approach, although the simple matrix in Eq. (\ref{TVxd}) can be applied, we choose to use the bicubic downsampling matrix, as it leads to better LR image. As discussed before, Cond. \ref{Cond_Fam} still holds in this case. Given the bicubic downsampling matrix, we test the repetition upsampling matrix in Eq. (\ref{EquRepetition}) as well as the bicubic upsampling matrix. It can be verified that Cond. \ref{Cond1} ${\bD_d} {\bU_d} = {\bI}$ holds approximately between these two upsampling matrices and the bicubic downsampling matrix.

{ In this paper, we use the Peak SNR (PSNR) to measure the objective quality of a reconstructed image, which is defined as $10{\log _{10}}({255^2}/{\rm{MSE}}({\bX} - {\hat \bX}))$, where ${\bX}$ is the reference image, and ${\hat \bX}$ is the test image.}

\subsubsection{Scaling Matrix $\bf{\Lambda }$}

During the reconstruction of LR image, in order to ensure that Cond. \ref{Cond_Qual} in Sec. \ref{PTC} is satisfied, we need to scale its corresponding measurement matrix ${\bA \bU_d}$ into ${\bA_d} = {\bA\bU_d\bf{\Lambda}} $ to get normalized columns, as shown in Eq. (\ref{ScaleTrans}) and Eq. (\ref{ScaleSpat}). Since no specific entries in the target LR image are preferred, the scaling matrix $\bf{\Lambda}$ should be diagonal matrix with equal diagonal entries.
For LR-AMP-ST in DCT and wavelet domain, the diagonal entry is the inverse of the downsampling factor $d$, according to Eq. (\ref{ScaleTrans}). For LR-AMP-TV-2D in TV domain, things are slightly different. For the repetition operator that replaces each pixel in LR image with a $d \times d$ block of pixels in the HR image, the diagonal entry in the scaling matrix is still $1/d$. For bicubic interpolation, we empirically set the diagonal entry in the scaling matrix to be $1/2.68$ for $d=2$ and $1/5$ for $d=4$. Although this approach cannot exactly normalize the columns and there are still some correlations between entries in the new measurement matrix, it works quite well in practices.

\subsubsection{Noiseless image recovery}

Tables \ref{DCTNoiseless}, \ref{WaveletNoiseless} and \ref{TVNoiseless} compare the  performances of DCT-domain MR-AMP-ST, wavelet-domain MR-AMP-ST, and spatial-domain MR-AMP-TV when there is no measurement noise. In each case, we compare our proposed LR-AMP that recovers the LR image directly, the conventional HR-AMP that reconstructs the HR image, and the naive H2L-AMP that recovers the HR image first and then downsamples it to obtain the LR image with the corresponding downsampling matrix. The highest PSNR in each case is highlighted.

From Tables \ref{DCTNoiseless} and \ref{WaveletNoiseless}, we can see that LR-AMP-ST almost always outperforms the other two algorithms, except when $d=4$ for HuskerStadium and SeaWorld. This is partially due to two reasons. First, land remote sensing images contain more details compared to natural images. Second, the suboptimal thresholding rule in \cite{DAMP45} is used for $d=4$, whereas the optimal SURE-based thresholding method in \cite{Paraless} is used for $d=2$.

In the spatial-domain approach, LR-AMP-TV-2D-B and H2L-AMP-TV-2D are the top two algorithms. Their reconstruction performances are comparable and the PSNR difference between them is within 1 dB. However, H2L-AMP-TV-2D is much slower than the proposed LR-AMP-TV-2D, as detailed in the computational complexity part later. Since the reference HR image is the same for the three approaches listed in Tables \ref{DCTNoiseless}, \ref{WaveletNoiseless} and \ref{TVNoiseless}, it can be seen that the TV-based approach yields higher PSNR than the transform-domain ones.

Fig. \ref{Visual} and Fig. \ref{HuskerStadium} illustrate the visual quality of the recovered $256 \times 256$  Barbara and Stadium by different methods. It can be seen that transform-domain and spatial-domain approaches have different types of reconstruction artifacts. The former preserves more details but also contains more high frequency noises, whereas the latter is blockier, despite higher PSNRs.

\subsubsection{Comparison between AMP-TV-2D-B and optimal TVAL3}
\label{AMPandTVAL3}

{In Table \ref{Compare1}, we compare the results of TVAL3 with optimized slack parameters \cite{Bell,TVAL3} and our parameter-free AMP-TV-2D-B for the MR-CS problem in Eq. (\ref{Model}). For the original HR image reconstruction, the performance of HR-AMP-TV-2D-B is comparable to the optimized HR-TVAL3. However, for the LR image reconstruction, our LR-AMP-TV-2D outperforms the optimized LR-TVAL3 in almost all cases by up to 1dB. More importantly, the theoretical analyses developed in Sec. \ref{PTC} and \ref{MRImage} are applicable for MR-AMP-TV, whereas there are only some qualitative analyses in \cite{Bell}.}

\subsubsection{Imaging in the presence of measurement noise}

Table \ref{NoisyResults} shows the performance of MR-AMP in different domains when various amounts of measurement noises are added. The proposed LR-AMP still outperforms the HR-AMP and H2L-AMP in almost all cases.

\subsubsection{LR approximation}
\label{secL2H}

{Another important problem in MR-CS is how to use a recovered LR image by LR-AMP to help the reconstruction of a higher-resolution image. As an initial attempt, we show in Table \ref{L2H} some results by simply upsampling the recovered LR image with the upsampling matrix to get a HR image, named L2H-AMP. As shown by the table, even this simple method can sometimes provide better HR images than HR-AMP. For example, L2H-AMP-ST can outperform HR-MP-ST in almost all cases. However, HR-AMP-TV-2D-B outperforms L2H-AMP-TV-2D-B when $d = 4$ and ${\delta _1} = 0.05 $, which implies that L2H-AMP is far from optimal. The reason is that high frequency information can be captured in CS measurements $\by$, but L2H-AMP is based on LR-AMP. It thus treats the high frequency information as approximation errors, and the upsampling matrix cannot estimate such information from LR image.}

\subsubsection{Computational complexity}

The computational complexities of various methods are reported in Table \ref{RunTime}, which shows that when $d=2$, the proposed LR-AMP is about $2$ times faster than the HR-AMP (the H2L-AMP is even slower than HR-AMP due to the additional downsampling), and the spatial-domain method is faster than the transform-domain one. However, when $d=4$ (the size of the LR image is $1/16$ of the HR one), the thresholding rule in soft-thresholding denoiser is changed from the time-consuming optimal SURE method in \cite{Paraless} for $d=2$ to the fast suboptimal max-min method in \cite{DAMP45}. Thus, the LR-AMP-ST is about $36$ times faster than HR-AMP-ST, the latter is about $25$ times faster than the HR-AMP-TV, and LR-AMP-ST is about $100$ times faster than LR-AMP-TV. Moreover, LR-AMP-TV is about $13$ times faster than HR-AMP-TV. This gives some guidelines on how to choose the appropriate method according to the value of $d$ when the complexity is a primary concern.

\section{Conclusion and future work}

In this paper, we systematically study the multi-resolution compressed sensing reconstruction problem, which can stably recover a low-resolution signal when the sampling rate is too low to recover the full resolution signal. We develop an AMP-based solution and study its theoretical performance. We also develop the appropriate up-/down-sampling operators in both transform and spatial domain. The performance of the proposed scheme is demonstrated via simulation results.

The proposed scheme can be further improved or applied to other applications. For example, in \cite{NCSUAMP, DAMP}, the authors introduce various latest image denoising algorithms into AMP. Better performance can be achieved if proper up-/down-sampling matrices can be designed for these denoisers. Another topic is to make full use of the LR-AMP to reconstruct better HR image, {\it i.e.}, to improve the performance of the L2H-AMP in Sec. \ref{secL2H}. Moreover, the proposed MR-AMP framework can also be applied to videos and multi-view images and videos \cite{MSCS}.


\appendix{ }
\label{PiecewiseProof}
\begin{appendices}

{

In this appendix, we prove that the condition in Corollary \ref{Concave}, {\it i.e.}, $M({\varepsilon _1}|\eta )$ is a concave function of ${\varepsilon _1}$, holds for the piecewise constant family in Eq. (\ref{Piecewise}).

We start by defining a special family of distributions for simple sparse signals:
\begin{equation}
\label{ssb}
{\mathcal{F}^{SS*}_{{n_1},{\varepsilon_1}}} \equiv \left\{ {\upsilon _{n_1}:
{\mathbb{E}_{{\upsilon _{n_1}}}} \left\{ {{{\left\| \bs{[2:n_1]} \right\|}_0}} \right\} \leqslant {n_1}{\varepsilon_1} } \right\},
\end{equation}
where $\bs[2:n_1]$ refers to the subvector of a signal $\bs$ from the second entry to the last entry.

Consider a signal $\bx$ in the piecewise constant signal family ${\mathcal{F}^{PC}_{{n_1},{\varepsilon_1}}}$ in Eq. (\ref{Piecewise}), and define $\bs$ as follows.
\begin{equation*}
\bs = {\left[ {x[1],{\rm{ }}x[2] - x[1],\dots, x[{n_1}] - x[{n_1} - 1]} \right]^T}.
\end{equation*}
It is clear that ${\bf{s}}\sim{v_{{n_1}}}$ where ${v_{{n_1}}} \in F_{{n_1},{\varepsilon _1}}^{SS*}$. Therefore a bijection relationship holds between ${\cal F}_{{n_1},{\varepsilon _1}}^{PC}$ and ${\cal F}_{{n_1},{\varepsilon _1}}^{SS*}$, since every signal generated from a distribution in ${\cal F}_{{n_1},{\varepsilon _1}}^{PC}$ is paired with exactly one signal from ${\cal F}_{{n_1},{\varepsilon _1}}^{SS*}$, and every signal from ${\cal F}_{{n_1},{\varepsilon _1}}^{SS*}$ is paired with exactly one signal from ${\cal F}_{{n_1},{\varepsilon _1}}^{PC}$. As a result, the proof in \cite{PTCPrediction} for the concavity of $M({\varepsilon _1}|\eta )$ for block-sparse signals is applicable to the piecewise constant family. However, the proof in \cite{PTCPrediction} (at the end of Page 3406) was very brief. Therefore we include the following details for completeness.

The goal of the concavity proof is to show that
\begin{equation}
\label{concave}
M( q \varepsilon_1 + (1-q)  \varepsilon_2 |\eta) \ge q M(\varepsilon_1 |\eta) + (1-q) M(\varepsilon_2 |\eta).
\end{equation}
First, from Eq. (\ref{SimpleSparse}) and (\ref{Piecewise}), if a distribution $\upsilon_{1} \in {\cal F}_{{n_1},q{\varepsilon _1} + (1 - q){\varepsilon _2}}$, then we have $\upsilon_{1} = q \upsilon_{2} + (1-q) \upsilon_{3}$, where $\upsilon_{2} \in {\cal F}_{{n_1},{\varepsilon _1}}$ and $\upsilon_{3} \in {\cal F}_{{n_1},{\varepsilon _2}}$, because any measure in ${\cal F}_{{n_1},q{\varepsilon _1} + (1 - q){\varepsilon _2}}$ can be written as a convex combination of measures in ${\cal F}_{{n_1},{\varepsilon _1}}$ and measures in ${\cal F}_{{n_1},{\varepsilon _2}}$\cite{PTCPrediction}. Next, note that $M({\varepsilon _1}|\eta )$ in Eq. (\ref{MinimaxMSE}) is obtained by tuning the denoising parameters to minimize the MSE of the least favorable distribution in the family. Eq. (\ref{concave}) can be proved by combining the two facts, because each term in the right hand side can be tuned independently to minimize its own least favorable MSE, whereas there is only one set of tuning parameters in the left hand side, leading to larger minimax MSE.
}

\end{appendices}

\section*{Acknowledgments}

The authors thank the reviewers for their valuable suggestions that have significantly enhanced the quality and presentation of the paper.

 \bibliographystyle{IEEEbib}
 \bibliography{refs}
\end{document}